\documentclass{aa}  
\usepackage{graphicx}
\usepackage{txfonts}
\usepackage{natbib}
\usepackage{subfigure}
\def\axj{AX~J1749.1$-$2733}
\def\integral{\emph{INTEGRAL}}
\def\xmm{\emph{XMM-Newton}}
\def\asca{\emph{ASCA}}
\def\swift{\emph{Swift}}
\def\chandra{\emph{Chandra}}
\def\gray{$\gamma$-ray}
\def\cps{counts$\,\mathrm{s}^{-1}$}
\newcommand{\unit}[2]{\mathrm{#1}^{#2}}
\def\ecms{\mathrm{ergs}\,\unit{cm}{-2}\,\unit{s}{-1}}
\def\es{\mathrm{ergs}\,\unit{s}{-1}}
\def\Ms{\mathrm{M}_{\odot}}

\def\cnu{$\chi_{\nu}^{2}$}
\def\mum{\mu\mathrm{m}}
\def\Ks{$K_{\mathrm{s}}$}
\newcommand{\seefig}[1]{(see Fig.~\ref{#1})}
\newcommand{\seetab}[1]{(see Table.~\ref{#1})}
\def\nh{N_{\mathrm{H}}}
\def\Av{A_{\mathrm{V}}}
\def\Ec{E_{\mathrm{cut}}}
\def\CI{C_{\mathrm{ISGRI}}}
\newcommand\ra[3]{#1^{\mathrm{h}}#2^{\mathrm{m}}#3^{\mathrm{s}}}
\newcommand\dec[3]{#1\degr#2\arcmin#3\arcsec}
\begin{document}
   \title{\integral, \xmm\ and ESO/NTT identification of \axj: an obscured and probably distant Be/X-ray binary\thanks{Based on observations made with 1) \integral, an ESA project with instruments and science data centre funded by ESA member states (especially the PI countries: Denmark, France, Germany, Italy, Switzerland, Spain), Czech Republic and Poland, and with the participation of Russia and the USA; 2) \xmm, an ESA science mission with instruments and contributions directly funded by ESA Member States and NASA; and 3) ESO Telescopes at the La Silla or Paranal Observatories under programme ID 079.D$-$0432(A).}}

   \author{J.A.~Zurita Heras
          \inst{1}
          \and
          S.~Chaty\inst{1}
          }

   \offprints{J.A.~Zurita Heras}

   \institute{Laboratoire AIM, CEA/DSM-CNRS-Universit\'e Paris Diderot,
	IRFU/Service d'Astrophysique, B\^at. 709, CEA-Saclay, FR-91191 Gif-sur-Yvette, France
	\email{juan-antonio.zurita-heras@cea.fr; sylvain.chaty@cea.fr}
   }

   \date{Received 19 November 2007; accepted 18 June 2008}

 
  \abstract
   {\axj\ is an unclassified transient X-ray source discovered during surveys by \asca\ in 1993--1999. The transient behaviour and the short and bright flares of the source have led to the idea that it is part of the recently revealed subclass of supergiant fast X-ray transients.}
   {A multi-wavelength study in NIR, optical, X-rays, and hard X-rays of \axj\ is undertaken in order to determine its nature.}
   {Public \integral\ data and our target of opportunity observation with \xmm\ were used to study the high-energy source through timing and spectral analysis. Multi-wavelength observations in optical and NIR with the ESO/NTT telescope were also performed to search for the counterpart.}
   {\axj\ is a new high-mass X-ray binary pulsar with an orbital period of $185.5\pm1.1$~days (or $185.5/f$ with $f=2,3\ \mathrm{or}\ 4$) and a spin period of $\sim 66$~s, parameters typical of a Be/X-ray binary. The outbursts last $\sim12$~d. A spin-down of $\dot{P}=0.08\pm0.02\ \mathrm{s}\,\unit{yr}{-1}$ is also observed, very likely due to the propeller effect. The most accurate X-ray position is R.A.~(2000)~$=\ra{17}{49}{06.8}$ and Dec.~$=\dec{-27}{32}{32.5}\ (\mathrm{uncertainty}\ 2\arcsec)$. The high-energy broad-band spectrum is well-fitted with an absorbed powerlaw and a high-energy cutoff with values $\nh=20.1_{-1.3}^{+1.5}\times10^{22}\ \unit{cm}{-2}$, $\Gamma=1.0_{-0.3}^{+0.1}$, and $\Ec=21_{-3}^{+5}$~keV. The only optical/NIR candidate counterpart within the X-ray error circle has magnitudes of $R=21.9\pm0.1$, $I=20.92\pm0.09$, $J=17.42\pm0.03$, $H=16.71\pm0.02$, and $K_{\mathrm{s}}=15.75\pm0.07$, which points towards a Be star located far away ($> 8.5$~kpc) and highly absorbed ($\nh\sim 1.7\times10^{22}\ \unit{cm}{-2}$). The average 22--50~keV luminosity is $0.4-0.9\times10^{36}\ \es$ during the long outbursts and $3\times10^{36}\ \es$ during the bright flare that occurred on MJD~52891 for an assumed distance of 8.5~kpc.}
   {}

   \keywords{X-ray: binaries -- X-rays: individual: \axj\ -- Infrared: stars
               }

   \authorrunning{Zurita \& Chaty}
   \titlerunning{\axj, an obscured and likely-distant Be/X-ray binary}
   \titlerunning{Multi-wavelength identification of \axj}
   \maketitle
%

\section{Introduction}

High-mass X-ray binaries (HMXB) are systems composed of a compact object (either a neutron star (NS) or a black hole (BH)) and a massive companion star. The emission in X-rays is explained by the accretion of the stellar wind coming from the companion and captured by the compact  object. There are mainly two classes of HMXB, depending on the nature of the massive primary star, which has important implications for the X-ray observational features. One might read \citet{Whiteal95} and \citet{Psaltis06} (and references therein) for reviews of HMXB.

The first class is composed of persistent high-energy binary systems made of a supergiant star (spectral type OB) (later called SGXB) orbiting close to the compact object.
The other group is composed of a Be star accompanying the compact object, mostly an NS. A Be star is a fast-rotating B star possessing a dense and slow stellar outflow at the equator (thus creating a disc-like ring of matter surrounding the B star). The group of BeXB is the most numerous in the HMXB family \citep{Liual00, Liual06}. They possess wide orbits ($P_{\mathrm{orb}}>15$~days). They display either recurrent outbursts near the periastron passage whose duration is normally linked to the orbital period (most common outbursts), or giant outbursts that can last from weeks to months. In most cases, their X-ray emission is modulated by the strong magnetic field of the NS. Their spectra are similar to those of supergiant X-ray binaries \citep[see reviews of][]{Bildstenal97,Coe00,Negueruela07}. 

Among the new sources, \integral\ might have unveiled a new subclass of SGXB. Indeed, several newly discovered sources were identified as galactic X-ray sources with a supergiant companion, and they display transient behaviour \citep[e.g. IGR~J17544$-$2619,][]{Zand05, Pellizzaal06}. The quiescent level in these systems is of the order of $10^{32-33}$~ergs$\,\mathrm{s}^{-1}$ \citep[e.g. IGR~J08408$-$4503,][]{Gotzal07,Leyderal07}. The X-ray luminosity increases up to $10^{36}$~ergs$\,\mathrm{s}^{-1}$ (as observed in the known SGXB) only during short luminous-flare activity. The X-ray luminosity remains at a very low level (if not totally absent) during most of the time, except during the flares. These flaring periods last for a few hours at most, and then the source goes back to an undetectable level of emission \citep[e.g. XTE~J1739$-$302,][]{Smithal06}. Therefore, they received the name of Supergiant Fast X-ray Transient (SFXT) \citep{Negueruelaal06}. Except for the transient nature of SFXT, their spectral features are similar to the previously known persistent SGXB.

The unclassified transient X-ray source AX J1749.1-2733 was discovered at a very low-luminosity level during the ASCA galactic centre survey performed between 1993 and 1999 \citep{Sakanoal02}. Its position is R.A.~(2000)~$=\ra{17}{49}{09.0}$ and Dec.~$=\dec{-27}{33}{13.9}$ (50$\arcsec$ at 90\% confidence). The source was detected three times over the six observations when the source was located within the instrument field of view. The observed 0.7--10~keV flux varies between $(2-6)\times 10^{-12}\ \ecms$. Each spectrum is fitted with an absorbed power-law model. However, only in one observation are the parameters well-constrained with $\Gamma=2.1_{-2.6}^{+4.7}$ and $\nh=25_{-21}^{+57}\times10^{22}\ \unit{cm}{-2}$.

A new flare of this source was observed by INTEGRAL on Sept. 9, 2003, lasting 1.3~days \citep{Sgueraal06}. The 20--60~keV peak flux was 40~mCrab, and they fitted the spectrum with a combination of a black body and a power-law ($KT=0.7_{-0.1}^{+0.3}\ \mathrm{keV}$ and $\Gamma=2.5\pm0.2$). Due to the shortness of the flare, they classified this source as a candidate SFXT. This tentative classification has already been pointed out by \citet{Zand05} and references therein. \citet{Grebeneval07} discussed one possible mechanism that could trigger such bright and fast flare in this system. If the source is a supergiant wind-fed accreting pulsar, the propeller mechanism would inhibit the accretion when the neutron star's spin period is less than the critical value $P_{\mathrm{s}}^{*}\sim3\,\mathrm{s}$. However, an increase in the local stellar wind density of the order of $(P_{\mathrm{s}}^{*}/P_{\mathrm{s}})^{7/3}$ during a change in the stellar outflow would result in the observed flare. \citet{Kuulkersal07} also report \integral\ detections in the periods Feb.--Apr. in 2005 and 2006 and non detection in 2005 Aug.--Oct.  with an average 20--60~keV flux of $<2$~mCrab in both detections\footnote{see the public page of the \integral\ Galactic Bulge program at {\tt http://isdc.unige.ch/Science/BULGE} for further information.}. Analysing all \integral\ public data and using the data available on the public page of the \integral\ Galactic Bulge program$^{1}$, \citet{Zuritaal07} discovered a likely period of $\sim185$~days as the source was detected by \integral\ at a level of $\gtrsim 5 \sigma$ for a few days  during March and September each year between 2003--2007.

Following this announcement, \swift\ performed a target of opportunity (ToO) observation of 5~ks on March 30, 2007 \citep{Romanoal07a}\footnote{From the same observation, \citet{Kong07} also report a slightly more accurate position with R.A.~(2000)~$=\ra{17}{49}{06.8}$ and Dec.~$=\dec{-27}{32}{31.5}$ (3.8$\arcsec$ at 90\% confidence). He also fitted the spectrum with an absorbed power-law with $\nh=(19\pm9)\times10^{22}\ \unit{cm}{-2}$, $\Gamma=2.1\pm1.2$, and an absorbed 2--10~keV flux of $\sim 6\times 10^{-12}\ \ecms$.}. They detected a bright source at the position R.A.~(2000)~$=\ra{17}{49}{06.8}$ and Dec.~$=\dec{-27}{32}{30.6}$ (6.3$\arcsec$ at 90\% confidence), $51\arcsec$ away from the \asca\ position\footnote{The distance was wrongly given in the ATel as their \asca\ position does not correspond to the one published in \citet{Sakanoal02}.}. Its spectrum could be fitted with an absorbed power-law ($\nh=23_{-10}^{+14}\times10^{22}\ \unit{cm}{-2}$, $\Gamma=2.5_{-1.7}^{+2.0}$) and the observed  2--10~keV flux is $\sim10^{-11}\ \ecms$. The \swift/UVOT telescope did not detect any optical counterpart with a $3\sigma$ upper limit of $V=20.67$~mag. Within the error circle, \citet{Romanoal07a} report three 2MASS candidate counterparts whose infrared (IR) magnitudes in the JHK bands suggest a strong optical extinction of 20~mag. Only one candidate is compatible with a supergiant. Thus, the nature of \axj\ remains a mystery.

\xmm\ also performed a ToO observation of the source during the expected outburst of 2007 March. From this observation, \citet{Karaseval07,Karaseval08} reported the discovery of a pulsation of $\sim132$~s. The pulse profile displays a double-peak structure with a pulse fraction of $\sim30$\% in the 2--10~keV energy range. They also detected a pulsation during the outburst detected by \integral\ on Sept. 8--10, 2003 in the 20--60~keV energy range with a higher pulse fraction of 50\%.

Here we report multiwavelength observations performed on \axj\ with \integral, \xmm, and the ESO/NTT telescope. In Sect.~\ref{secObs}, we first describe the observations and the data analysis of each instrument. Then, we present the results in Sect.~\ref{secRes}. Finally, we finish with a discussion and the conclusion on the nature of the source in Sect.~\ref{secDis}.

\section{Observations and data analysis}\label{secObs}
The present work is based on data of two high-energy space missions, \integral\ \citep{Winkleral03} and \xmm\ \citep{Jansenal01}, of the European Space Agency (ESA). Multi-wavelength follow-up observations were also performed with the ESO\footnote{European Southern Observatory} 3.5~m New Technology Telescope (NTT) at La Silla Observatory, Chile.

\subsection{\integral}

The INTErnational Gamma-Ray Astrophysics Laboratory (\integral) is a hard X-ray and \gray\ spacecraft (S/C) laboratory operating since Oct. 2002. The scientific payload is composed of four instruments. However, only data from the hard X-ray and soft \gray\ coded-mask imager IBIS/ISGRI (15~keV--1~MeV) \citep{Ubertinial03,Lebrunal03} are going to be considered in this work. The imager possesses a wide field of view (FOV) of 29$\degr$ square with a spatial resolution of 12$\arcmin$. 

The source is located near the galactic centre at a distance of 1.6$\degr$. As the galactic centre is one of the major scientific goals of \integral, the source's field has been extensively observed. All public data available in March 2007 for \axj\ have been considered in this work. Only pointings where the source is located less than 14$\degr$ from the FOV centre and whose time exposure is longer than 600~s are kept. In total, we collected 4759 pointings distributed in 129 revolutions of the S/C that goes from Feb. 2003 (revolution 46, MJD~52698.0) to Oct. 2005 (revolution 370, MJD~53670.1). The total exposure time on the source is 10.8~Ms ({\it i.e.} 125.4~days) spanning 2.5~years of observations. They are not equally distributed along this period for scheduling reasons.

The ISGRI data were reduced using the Offline Scientific Analysis (OSA\footnote{OSA is available at {\tt http://isdc.unige.ch/?Soft+download}}) version 6.0 software that is publicly released by the \integral\ Science Data Centre (ISDC) \citep{Courvoisieral03}. Individual sky images for each pointing were produced in the energy band 22--50~keV. Sky mosaics with longer exposures were built combining pointings in which the source is not detected at a 5.1$\sigma$ level or higher in individual pointings. The light curves were built using the imaging products. The source count rate was extracted with help of the tool {\tt mosaic\_spec} (version 1.4) that is part of the OSA package. Detections of the source  in mosaics are considered at a 6$\sigma$ level or higher. We extracted a spectrum of the source during the first bright flare detected with ISGRI. The spectral extraction was performed using the recently released OSA version 7.0. The source spectrum was extracted with {\tt ii\_spectra\_extract} within OSA for each pointing. Then, each individual spectra was summed to build one single  spectrum of the source using the OSA tool {\tt spe\_pick}. The redistribution matrix file (RMF) {\tt isgr\_rmf\_grp\_0025.fits} was rebinned into 5 channels spread between 15 and 80~keV. Light curves with a binning of 10~s of the first bright flare observed with ISGRI were also extracted with {\tt ii\_light} version 8.2. The light curves are given in counts$\,$s$^{-1}$. The 22--50~keV count rates can be converted into other units with the relation $1\,\mathrm{Crab}=117\,\mathrm{counts}\,\mathrm{s}^{-1}=9\times10^{-9}\ \ecms$. The time unit IJD corresponds to $\mathrm{MJD}=\mathrm{IJD}+51544$.

\subsection{\xmm}

The main scientific instrument onboard the X-ray Multi-Mirror Mission (\xmm) satellite is the EPIC camera composed of two MOS \citep{Turneral01} and one pn \citep{Struderal01} CCD cameras. It has imaging, timing, and spectral capabilities in the 0.2--12~keV energy range with a 30$\arcmin$ FOV. The EPIC cameras were operating in imaging science mode with full window for MOS1, large window for MOS2 and pn, and with medium filters for each camera.

 \axj\ was observed with \xmm\ on March 31, 2007, from 08:03:37 to 11:27:40 UTC (MJD~54190.337--54190.478) for a total exposure of 12.2~ks. There was no simultaneous observation with \integral.
 
Event lists for each MOS and pn camera were generated with the Science Analysis Software (SAS\footnote{SAS is available at {\tt http://xmm.vilspa.esa.es/external\\/xmm\_sw\_cal/sas.shtml}}) version 7.0.0 using the {\tt emproc} and {\tt epproc} tools, respectively. The event lists were corrected for enhanced background features observed at energies higher than 10~keV, disregarding time lapses when count rates exceeded 0.52~\cps\ for pn and 0.2~\cps\ for both MOS. Therefore, the net exposure is 5.6 out of 10~ks exposure for pn, 10.8 out of 12~ks for MOS1, and 11.1 out of 12~ks for MOS2. Images for MOS[12] and pn were generated with 2$\arcsec$ and 4$\arcsec$ resolution, respectively, using good events until the quadruple level in the 0.8--10~keV energy range and disregarding bad pixels. An accurate X-ray position determined with EPIC was calculated with the SAS task {\tt edetect\_chain}. Four images with energy ranges 0.5--2, 2--4.5, 4.5--7.5, and 7.5--12~keV for each MOS and pn camera were extracted. Then, the best position for each individual EPIC camera was determined. Finally, the best source position was calculated as the mean of the positions given by the three cameras. The source location precision is limited by the astrometry of the S/C, that is of 2$\arcsec$\footnote{The \xmm\ astrometric accuracy is described in the calibration document {\tt XMM-CAL-TN-0018}.}, the statistical error of $0.1\arcsec$ being insignificant in comparison.

In the EPIC/pn image, one bright source was detected in the CCD\#1 near the read-out node. The event list of the source$+$background was extracted from a region of 35$\arcsec$ radius centred on the source. A background event list was extracted in the same CCD at same distance of the read-out node from a region of similar size not affected by the bright source. Spectra were extracted by selecting single$+$double events but disregarding bad pixels. Specific RMF and ancillary response files (ARF) were generated with the SAS tasks {\tt rmfgen} and {\tt arfgen}, respectively. The {\tt Xspec} version 11.3.2t package \citep{Arnaud96} was used to plot and fit the resulting spectra corrected from background. For light curves, single and double events were also selected within the same regions defined in the spectral step. The source light curves were corrected from the background using the SAS task {\tt lccorr}, and we applied the barycentric correction with the SAS task {\tt barycen}.

\subsection{ESO/NTT}

Multiwavelength observations were undertaken soon after the accurate X-ray position was given by \citet{Romanoal07a} and \citet{Kong07}. The observations were achieved with the 3.5~m ESO/NTT telescope in two domains: in the near IR (NIR) domain ($1-2.4\ \mum$) with the spectro-imager SOFI and in the optical domain with the imager SUSI-2 ($350-900$~nm). The observations were carried out as part of the ToO programme 079.D-0432(A) (PI: S.~Chaty), through service mode. Astrometry, photometry, and spectroscopy were achieved during those observations. The spectroscopy was performed on the bright counterpart proposed by \citet{Romanoal07a} (2MASS source \#1).

\subsubsection{NIR observations}

The NIR photometry in the bands $J$, $H$, and $K_{\mathrm{S}}$ was performed on April 3, 2007, with the spectro-imager SOFI installed on a Nasmyth focus of the NTT. The observations were centred on the known X-ray position of \axj. The large imaging mode was used during those observations, and it gives an image pixel scale of $0.288\arcsec$ and an FOV of $4.92\arcmin\times 4.92\arcmin$. Nine images were taken for each band with integration times of 60~s for $J$ and 47.3~s for $H$ and \Ks. For each band, four of the nine observations were taken with a slight offset of $\sim30\arcsec$ that allowed us to build the NIR sky to subtract it from the images. Six standard stars chosen from the faint NIR standard star catalogue \citep{Perssonal98} were also observed: S262-E, S495-E, S708-D, S264-D, P565-C, and S875-C. Five observations per band were performed on each standard star. The first observation was centred on the target and then, the next four images were taken with an offset of $\sim45\arcsec$ compared to the first one.

\subsubsection{Optical observations}

The optical observations were carried out on April 6, 2007, between UTC 08h40 and 08h40 with the imager SUSI-2 also installed on the same Nasmyth focus of the NTT as SOFI. Optical photometry in $U$, $B$, $V$, $R$, $I$, and $Z$ bands was obtained. The FOV was $5.5\arcmin\times 5.5\arcmin$ with a binning of factor 2 that gives a pixel scale of $0.161\arcsec$ per pixel. The exposure time is 60~s for each filter. Nine photometric standards \citep[selected in the optical standard star catalogue of][]{Landolt92} were also observed in the fields PG~1657+078 and PG~1633+099. The integration times in each filter varied between 10--100~s.

\subsubsection{Data reduction}

The reduction of both optical and NIR data was performed with the Image Reduction and Analysis Facility (IRAF\footnote{IRAF is available at {\tt http://iraf.net/}}) version 2.13beta2. Data reduction was performed using standard procedures on the optical and NIR images, including the correction of the dark current, the flat-fielding, and NIR sky subtraction.

We performed accurate astrometry on each image ($U,B,V,R,I,Z,J,H,K_{\mathrm{S}}$) using the {\tt gaia} tool from the Starlink suite\footnote{Thanks to the Joint Astronomy Centre, the Starlink software collection is still available at {\tt http://starlink.jach.hawaii.edu/}.} and using the 2MASS catalogue for the NIR images and the USNO B1.0 catalogue for the optical images. The NIR rms of astrometry fit was always lower than $0.4\arcsec$ with the expected pixel scale in x,y axis of $-0.288\times0.288\arcsec$/pixel. The optical rms of astrometry fit was also lower than $0.4\arcsec$ with the expected pixel scale in x,y axis of $0.161\times0.161\arcsec$/pixel. 

Then, we carried out aperture photometry in a crowded field using the IRAF {\tt digiphot.daophot} package. The instrumental magnitudes $m_{\mathrm{instr}}$ were transformed into apparent magnitudes $m_{\mathrm{app}}$ using the standard photometric relation: $m_{\mathrm{app}}=m_{\mathrm{instr}}-Z_{\mathrm{p}}-ext\times AM$, where $Z_{\mathrm{p}}$ is the zero-point, $ext$ the extinction, and $AM$ the airmass. The colour term was not used because 1) there were not enough standard stars in NIR and 2) in optical bands, we did not detect the sources of interest in several bands (particularly in $V$) in order to perform colour correction. The $Z_{\mathrm{p}}$ and $ext$ parameters were fitted using the photometric relation in order to match the instrumental and apparent magnitudes of the standard stars. For the optical photometry, the extinction parameters were fixed to the values given during the SUSI-2 calibration\footnote{retrieved at {\tt http://www.ls.eso.org/lasilla/sciops/ntt/}\\ {\tt susi/archive/susi2zp.dat}}. The airmass, zero-point, and extinction parameters are reported at the end of Table~\ref{tab_photom}.

\section{Results}\label{secRes}

\subsection{Refined X-ray position}\label{secPos}

\begin{figure}
\centering
\resizebox{\hsize}{!}{\includegraphics{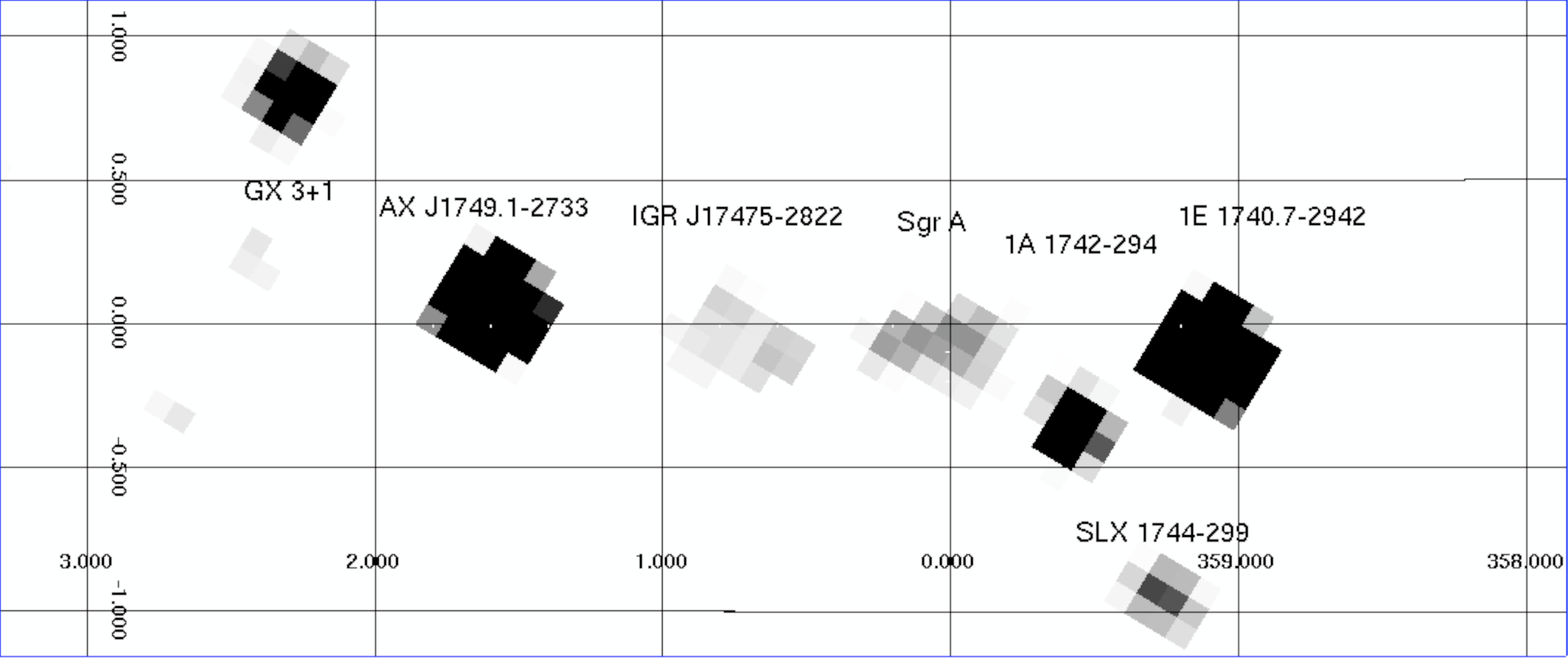}}
\caption{ISGRI 22--50~keV significance map built with the pointings of rev.~{\tt 110} listed in Table~\ref{tab_flare} corresponding to the first outburst of \axj\ observed with \integral. \axj\ is located near the galactic centre, and it is detected with a significance of 44$\sigma$ for an exposure of 51.3~ks. The coordinate grid corresponds to the galactic longitude and latitude $l,b$ in degrees.}
\label{isgri_image}
\end{figure}

A 22--50~keV image was generated from the first outburst whose pointings are listed in Table~\ref{tab_flare}. The source is detected at a level of 44$\sigma$ for a total exposure of 51.3~ks \seefig{isgri_image}.
The best  hard X-ray position is R.A.~(2000)~$=\ra{17}{49}{05.5}$ and Dec.~$=\dec{-27}{32}{38.4}$ with an error of 1.8$\arcmin$ related to its significance \citep{Grosal03}. This position is consistent with the position previously reported in the 3rd ISGRI catalogue \citep{Birdal07} and in \citet{Sgueraal06}.

\begin{figure}
\centering
\resizebox{\hsize}{!}{\includegraphics{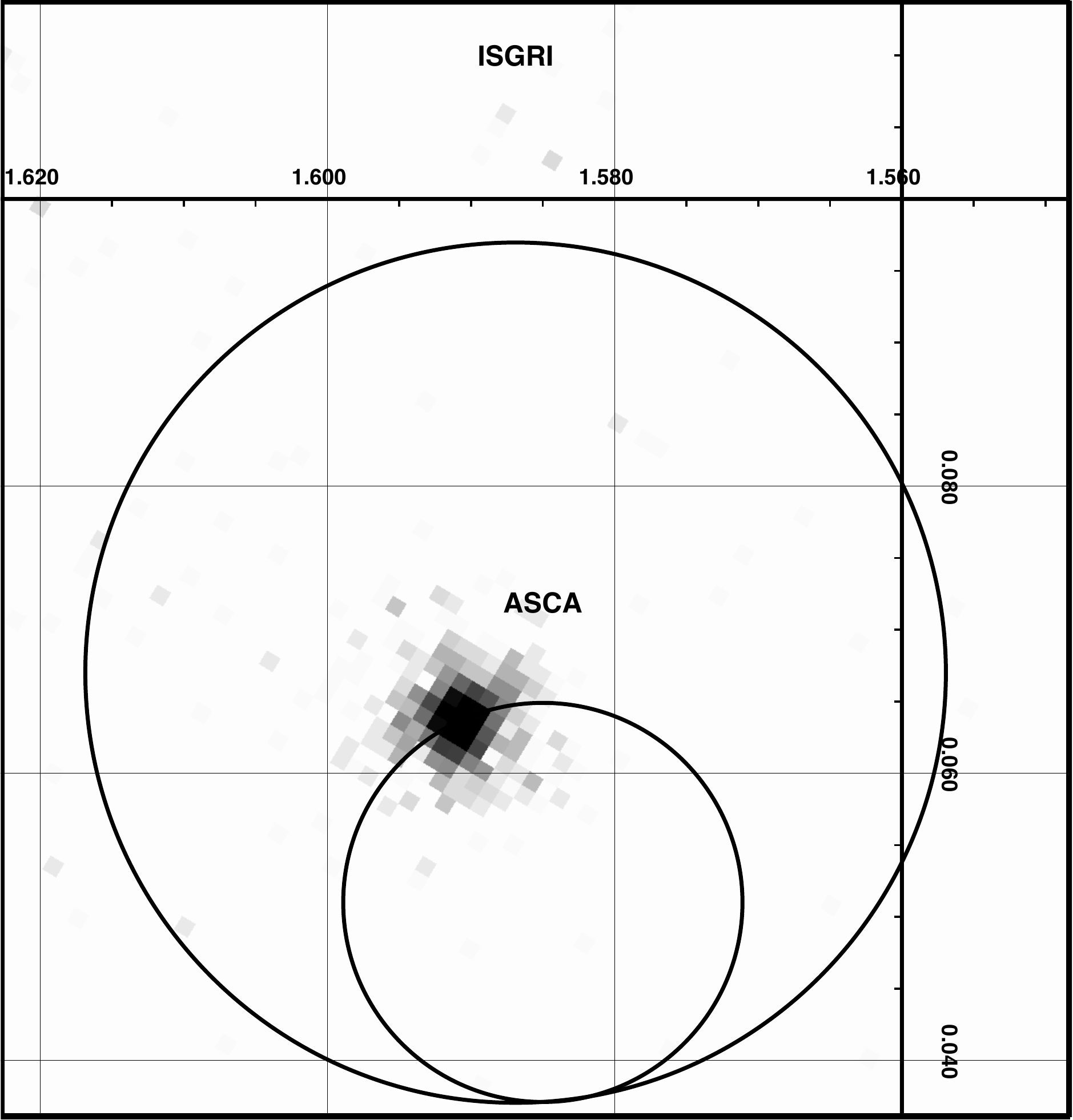}}
\caption{EPIC/MOS1 image in the 0.8--10 keV energy range. The ISGRI (derived from the first outburst) and ASCA error circles are reported, with the \swift\ position embedded in the bright source. The best EPIC position is R.A.~(2000)~$=\ra{17}{49}{06.8}$ and Dec.~$=\dec{-27}{32}{32.5}$ with a systematic uncertainty of 2$\arcsec$ (e.g. galactic $l,b=1.5908\degr,0.0637\degr$). The coordinate grid corresponds to the galactic longitude and latitude $l,b$ in degrees.}
\label{mos1_image}
\end{figure}

One bright source lies within the \axj\ ISGRI error circle in EPIC images \seefig{mos1_image}. The best position determined with EPIC is R.A.~(2000)~$=\ra{17}{49}{06.8}$ and Dec.~$=\dec{-27}{32}{32.5}$ with a systematic uncertainty of 2$\arcsec$. This X-ray refined position is consistent with the respective error circles of  \asca, \integral, and \swift. There is only a difference of $1\arcsec$ between the \swift\ and the EPIC positions.

\subsection{Timing analysis}

While the long-term variability of \axj\ can be studied with the huge amount of data collected with \integral, \xmm\ allows study of the short-term variability of the source.

\subsubsection{Long-term variability}

\begin{figure}
\centering
\resizebox{\hsize}{!}{\includegraphics{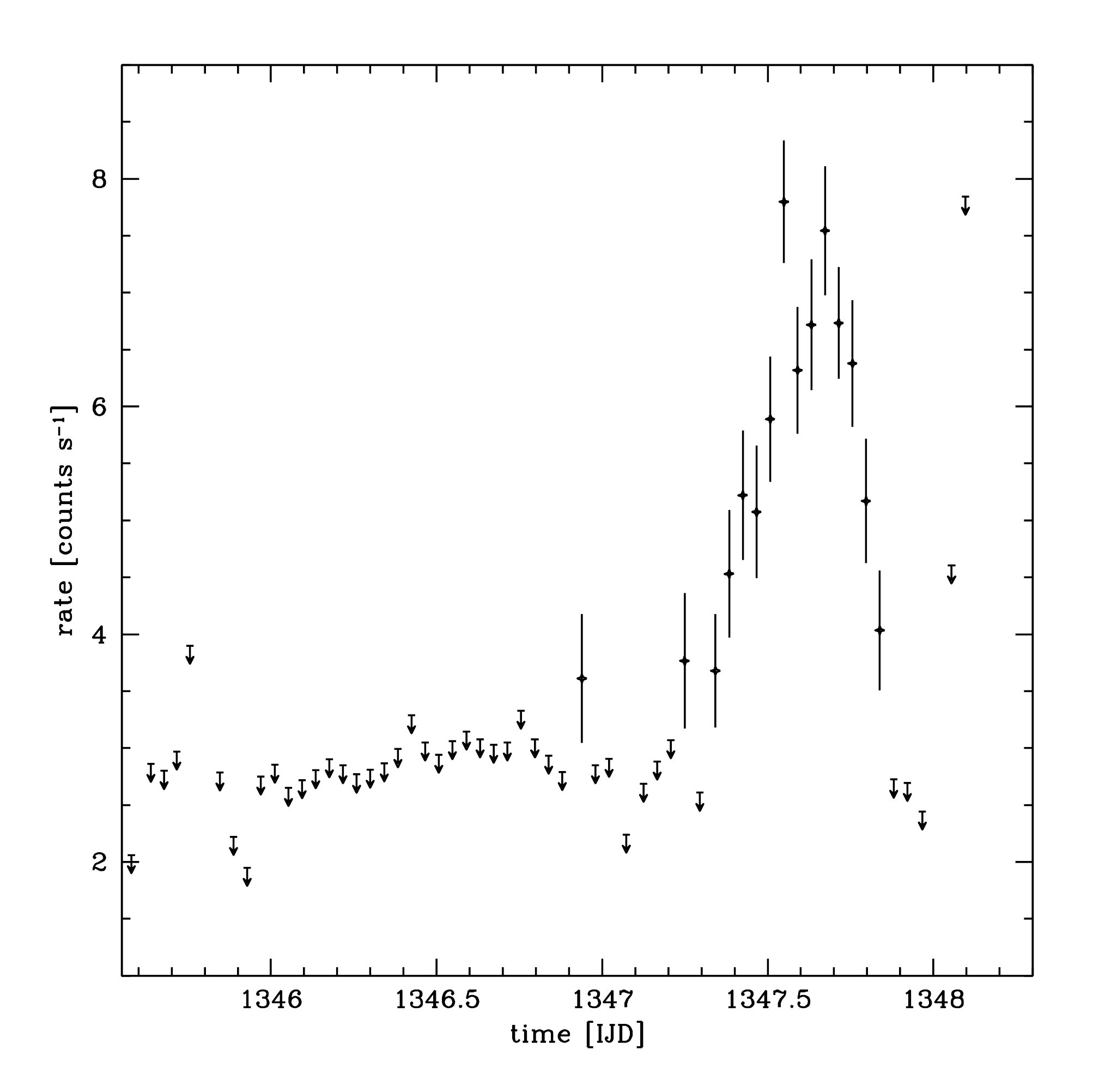}}
\caption{ISGRI 22--50~keV light curve of \axj\ during rev.~{\tt 110}. The 5.1$\sigma$ upper limits are reported when the source was not detected. }
\label{isgri_fl1}
\end{figure}

\begin{figure}
\centering
\resizebox{\hsize}{!}{\includegraphics{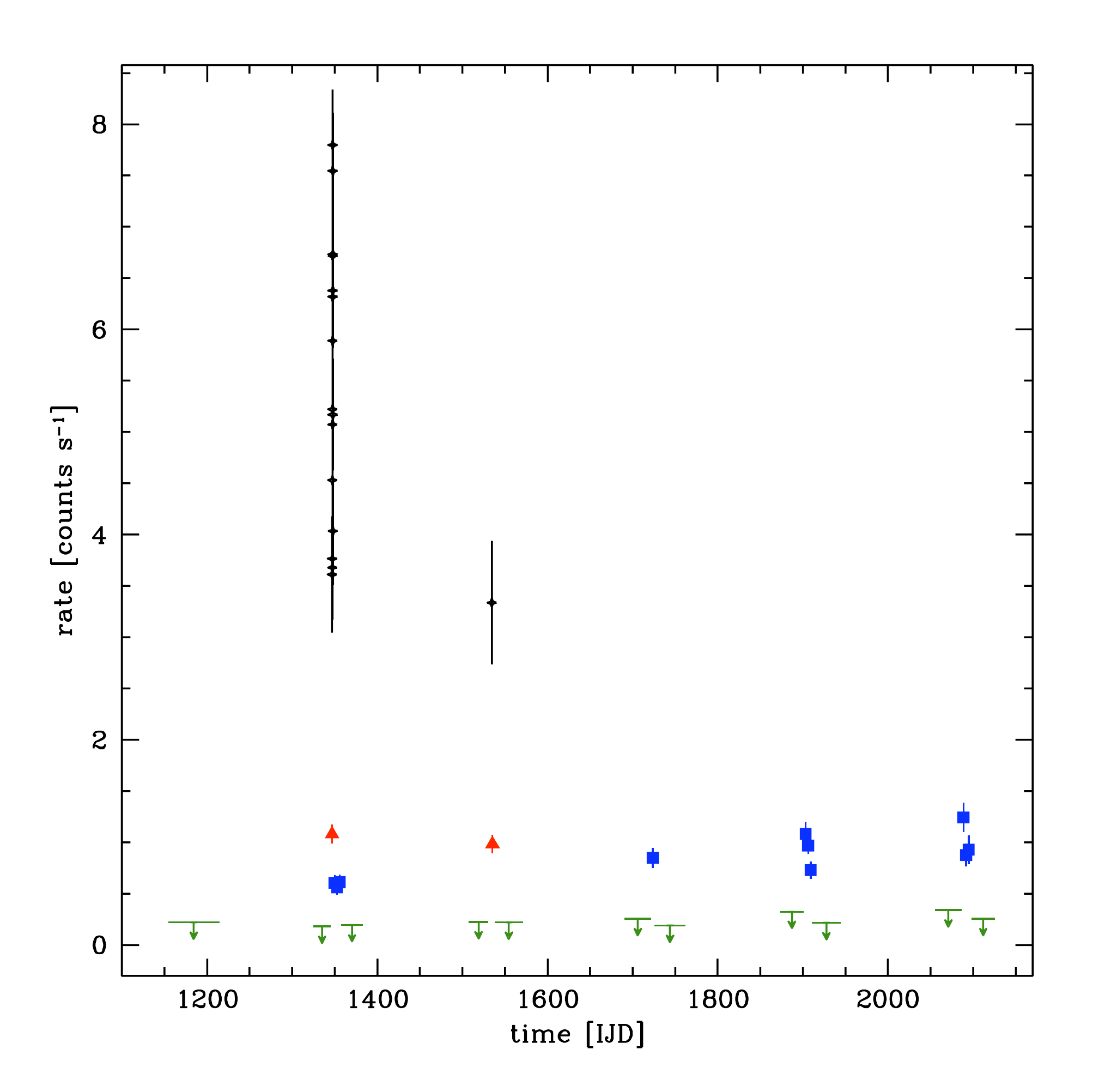}}
\caption{ISGRI 22--50~keV long-term light curve of \axj. The 2 flares reported in Table~\ref{tab_flare} are shown as vertical crosses for each single pointing. The other points correspond to average count rates extracted in large mosaic images. Details about symbols are found in the text.}
\label{isgri_lc}
\end{figure}

The long-term variability of \axj\ was studied using the ISGRI data in the energy range 22--50~keV.
All detections within one pointing of the S/C indicating an increase in the hard X-ray emission on a short time-scale ($\lesssim 1$~h) were searched. In total, the source was detected at a level higher than 5.1$\sigma$ in only 16 pointings out of the 4759 available \seetab{tab_flare}. Two bright flares were observed during the 2.5~yr observations, one lasting $\lesssim 1$~day in revolution {\tt 110}\footnote{This one is reported in \citet{Sgueraal06}.} \seefig{isgri_fl1} and another short one in revolution 173 lasting 1~h. In revolution {\tt 110}, the source was detected in two single pointings at a level slightly higher than the detection threshold before displaying a bright flare that started at IJD 1347.3636 and lasts $\sim0.5$~days. The count rate's increase/decrease was rather smooth. The peak intensity was reached in the middle of the outburst with two local maxima at $7.5-7.8\pm0.5$~\cps\ and a small decay in between. The two flares are also reported  in the long-term light curve \seefig{isgri_lc}. 
\begin{table*}
\caption{List of detections of \axj\ in single pointings with IBIS/ISGRI.}
\label{tab_flare}      
\centering          
\begin{tabular}{c c c c c c}
\hline\hline       
Time & 22--50~keV Flux & Error flux & Sigma & Off-axis angle & Pointing\\
IJD & \cps & \cps & ~ & degrees & ~ \\
\hline                    
1346.9390 & 3.6 & 0.6 & 6.4 & 3.7 & {\tt 01100038} \\
1347.2492 & 3.8 & 0.6 & 6.3 & 4.7 & {\tt 01100045} \\
1347.3421 & 3.7 & 0.5 & 7.4 & 0.8 & {\tt 01100047} \\
1347.3836 & 4.5 & 0.6 & 8.1 & 1.5 & {\tt 01100048} \\
1347.4250 & 5.2 & 0.6 & 9.2 & 3.4 & {\tt 01100049} \\
1347.4664 & 5.1 & 0.6 & 8.7 & 4.2 & {\tt 01100050}  \\
1347.5078 & 5.9 & 0.6 & 10.7 & 2.9 & {\tt 01100051} \\
1347.5491 & 7.8 & 0.5 & 14.5 & 2.6 & {\tt 01100052} \\
1347.5903 & 6.3 & 0.6 & 11.4 & 3.7 & {\tt 01100053} \\
1347.6318 & 6.7 & 0.6 & 11.7 & 5.3 & {\tt 01100054} \\
1347.6732 & 7.5 & 0.6 & 13.3 & 2.7 & {\tt 01100055} \\
1347.7146 & 6.7 & 0.5 & 13.7 & 0.8 & {\tt 01100056} \\
1347.7560 & 6.4 & 0.6 & 11.5 & 1.5 & {\tt 01100057} \\
1347.7972 & 5.2 & 0.5 & 9.5 & 2.0 & {\tt 01100058} \\
1347.8385 & 4.0 & 0.5 & 7.7 & 1.6 & {\tt 01100059} \\
1534.6049 & 3.3 & 0.6 & 5.5 & 4.9 & {\tt 01730024} \\
\hline
\end{tabular}
\end{table*}

Then, detections on a larger timescale are searched. The long-term monitoring of \axj\ is summarised in Table~\ref{tab_quiescent}. We built mosaics for each revolution with the source within the FOV. Concerning revolutions {\tt 110} and {\tt 173}, the pointings where \axj\ was detected were discarded before building both mosaics. The exposures are long with 127 and 144~ks, respectively. In both cases, the source was detected significantly with $11\sigma$ and a count rate of $1.0\pm0.1$ (triangle points in Fig.~\ref{isgri_lc}). In the other mosaics, the source was detected in 10 revolutions at a significant level of $>6\sigma$ for exposure times between 100--200~ks (square points in Fig.~\ref{isgri_lc}). 

All the detections are distributed in 5 outbursts that are separated by half a year each time. The two flares reported above occurred during outbursts 1 and 2. There are 2 isolated revolutions and three periods where the source is detected in consecutive revolutions (3 or more) giving elapsed time of activity of 11.6, 2.5, 2.3, 8.4, and 8.3 days. However, the observed flux level during the outbursts is faint with rates of $(0.56-1.25)\pm0.1$~\cps. With such faintness, the exposure time in individual revolution needs to be higher than $\sim 100$~ks in order to detect the source. For each observed outburst except the first one, the revolutions prior to and/or after the outburst do not fill this condition. Considering the nearest revolutions with an exposure time longer than 100~ks and surrounding the outbursts, the maximum outburst duration that we derive is $\lesssim18$~d. Only the first outburst duration can be constrained to less than 12.4~d, which is the difference between the ending and start of rev. {\tt 109} and {\tt 114}. Both revolution exposure times are $\sim 200$~ks, so it firmly excludes any activity by \axj\ with the same intensity displayed in the other detections. The outbursts $1,4,5$ all display similar behaviour with a first detection $\gtrsim 1$~count$\,\unit{s}{-1}$ followed by fainter detections. Considering the start of these 3 outbursts, we derive a period of $185.5\pm1.1$~d. The two shorter outbursts 2 and 3 are consistent with this period and an outburst's duration of  $\lesssim 12$ days. The source was not detected during the first part of the observation as the expected outburst time falls in the middle of a gap of 12~d without observation. Moreover, the period of \axj\ may be shorter by a factor $185.5/f$~d where $f=2,3,\ \mathrm{or}\ 4$ once considered the total unequal coverage of the source field by \integral\ over the years. However, \integral\ observations discard factors $f\geq5$. 

The \asca\ detections are 10 times fainter than the \xmm\ ones that occurred during an outburst of the source. \integral\ cannot reach this faintness, as detected by \asca, even for exposure of several Ms. They fall at phases 0.2--0.28 when the period is 185.5~d. If the source detected by \asca\ is the same one detected by both \integral\ and \xmm, either the outbursts last longer with a duration of $\sim50$~d, with \asca\ observing the end of the outburst, or the period of the system is shorter. In the last case, \asca\ detections would be compatible with $f=4$, making a period of 46.4~d.
\begin{table*}
\caption{Long-term monitoring of \axj\ with \integral/IBIS/ISGRI.}
\label{tab_quiescent}      
\centering          
\begin{tabular}{c c c c c r c}
\hline\hline       
Time start& Time stop & 22--50~keV Flux & Flux error & Detection & Exposure & revolution or\\
IJD & IJD & \cps & \cps & sigma & s & observed fraction$^\dagger$ \\
\hline                    
\multicolumn{6}{l}{\it one revolution} \\
1345.5410 & 1348.1174 & 1.08 & 0.10 & 11.4 & 126803 & {\tt 110}\\
1348.6103 & 1351.1086 & 0.61 & 0.08 & 7.8 & 176622 & {\tt 111} \\
1351.5239 & 1354.1001 & 0.56 & 0.07 & 7.8 & 203381 & {\tt 112} \\
1354.5150 & 1357.0919 & 0.61 & 0.07 & 8.3 & 195763 & {\tt 113} \\
1534.1522 & 1536.6650 & 0.99 & 0.09 & 10.8 & 143943 & {\tt 173} \\
1722.5068 & 1724.7996 & 0.85 & 0.10 & 8.8 & 147812 & {\tt 236} \\
1902.1512 & 1904.5188 & 1.08 & 0.12 & 9.2 & 137347 & {\tt 296} \\
1904.9797 & 1907.4973 & 0.97 & 0.08 & 11.9 & 192511 & {\tt 297} \\
1907.9482 & 1910.5226 & 0.73 & 0.09 & 8.6 & 177189 & {\tt 298} \\
2087.5924 & 2089.8853 & 1.25 & 0.15 & 8.6 & 101696 & {\tt 358} \\
2090.6097 & 2092.9742 & 0.88 & 0.11 & 8.0 & 129327 & {\tt 359} \\
2093.5752 & 2095.9587 & 0.93 & 0.14 & 6.5 & 108779 & {\tt 360} \\
\multicolumn{6}{l}{\it multi-revolutions} \\
1154.1922 & 1214.5034 & 0.34 & 0.04 & 7.9 & 887100 & 0.17\\
1324.9592 & 1345.1323 & 0.21 & 0.04 & 6.0 & 1004817 & 0.17\\
1357.5105 & 1383.3782 & 0.26 & 0.04 & 6.6 & 904838 & 0.40\\
1507.4945 & 1530.5140 & 0.31 & 0.05 & 7.0 & 687994 & 0.35\\
1537.7486 & 1571.5313 & 0.35 & 0.04 & 7.8 & 668850 & 0.23\\
1690.1832 & 1722.0688 & 0.36 & 0.05 & 7.0 & 603794 & 0.22\\
1725.8560 & 1762.0251 & 0.26 & 0.04 & 6.8 & 976316 & 0.31\\
1873.1647 & 1900.9215 & 0.37 & 0.07 & 5.8 & 333370 & 0.14\\
1910.9388 & 1944.1377 & 0.37 & 0.04 & 8.8 & 770675 & 0.27\\
2055.1532 & 2086.8640 & 0.42 & 0.07 & 6.2 & 389952 & 0.14\\
2097.9383 & 2125.8160 & 0.47 & 0.05 & 9.1 & 658682 & 0.27\\
\hline                  
\end{tabular}
\begin{list}{}{}
\item[$^\dagger$either the revolution number ({\it top}) or the observed fraction derived as exposure/(time stop-time start) ({\it bottom}).].
\end{list}
\end{table*}

The revolutions where the source is not detected are accumulated only when they are separated by less than 30 days. This results in 11 mosaics with exposure times lying between 333--1005~ks. The 1$\sigma$ level lies between 0.04--0.07~\cps ($=$0.05--0.08~mCrab). As the field is noisier with exposure times $>300$~ks, the detection limit is set to $10\sigma$ ({\it i.e.} $\gtrsim 0.5$~mCrab). The fluxes extracted in these mosaics are reported in the {\it multi-revolutions} part of Table~\ref{tab_quiescent} (upper limit arrows in Fig.~\ref{isgri_lc}, the observed fraction of these upper limits is reported in Table~\ref{tab_quiescent}). The source is never detected besides the 5 outbursts discussed earlier. The detailed 22--50~keV light curve is shown in Fig.~\ref{isgri_lc}.

\subsubsection{Short-term variability}

The short-term variability of \axj\ is studied using the EPIC data. The 0.4--10~keV light curve is rather constant with a low average count rate of $\sim0.4$~\cps. No flare is observed. Considering the period and duration of outbursts derived from the \integral\ long-term light curve, the expected outburst near the \xmm\ observation happens between MJD~54188 and 54206. Therefore, \xmm\ observed \axj\ at the beginning of one of its outburst (MJD~54190).

\begin{figure*}[htbp]
  \centering
  \mbox{
  \subfigure[EPIC/pn periodogram]{\includegraphics[width=0.5\textwidth]{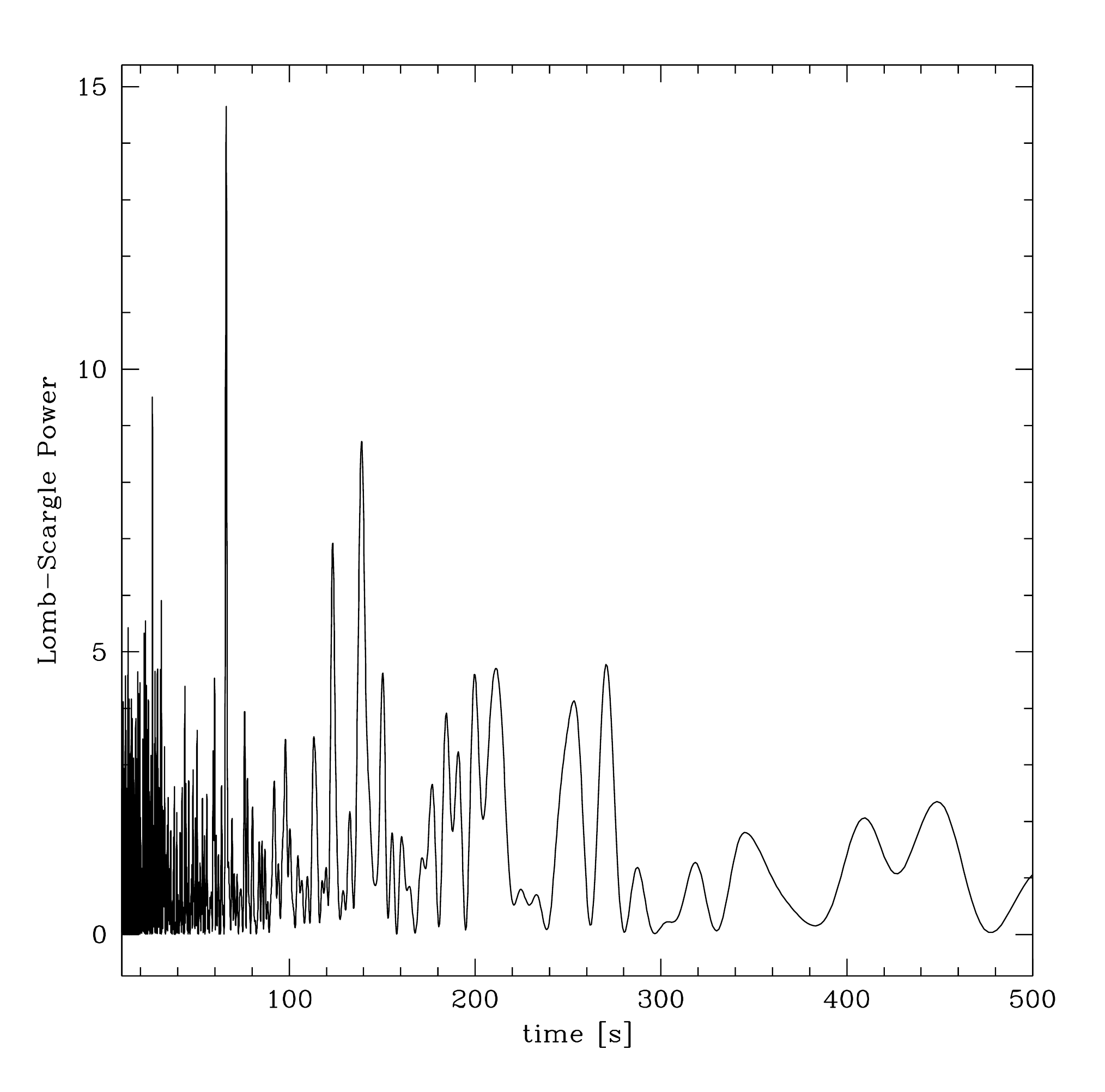}}
  \subfigure[pn folded light curve, $P=66.09\pm0.07$~s and T0=MJD 54190.391]{\includegraphics[width=0.5\textwidth]{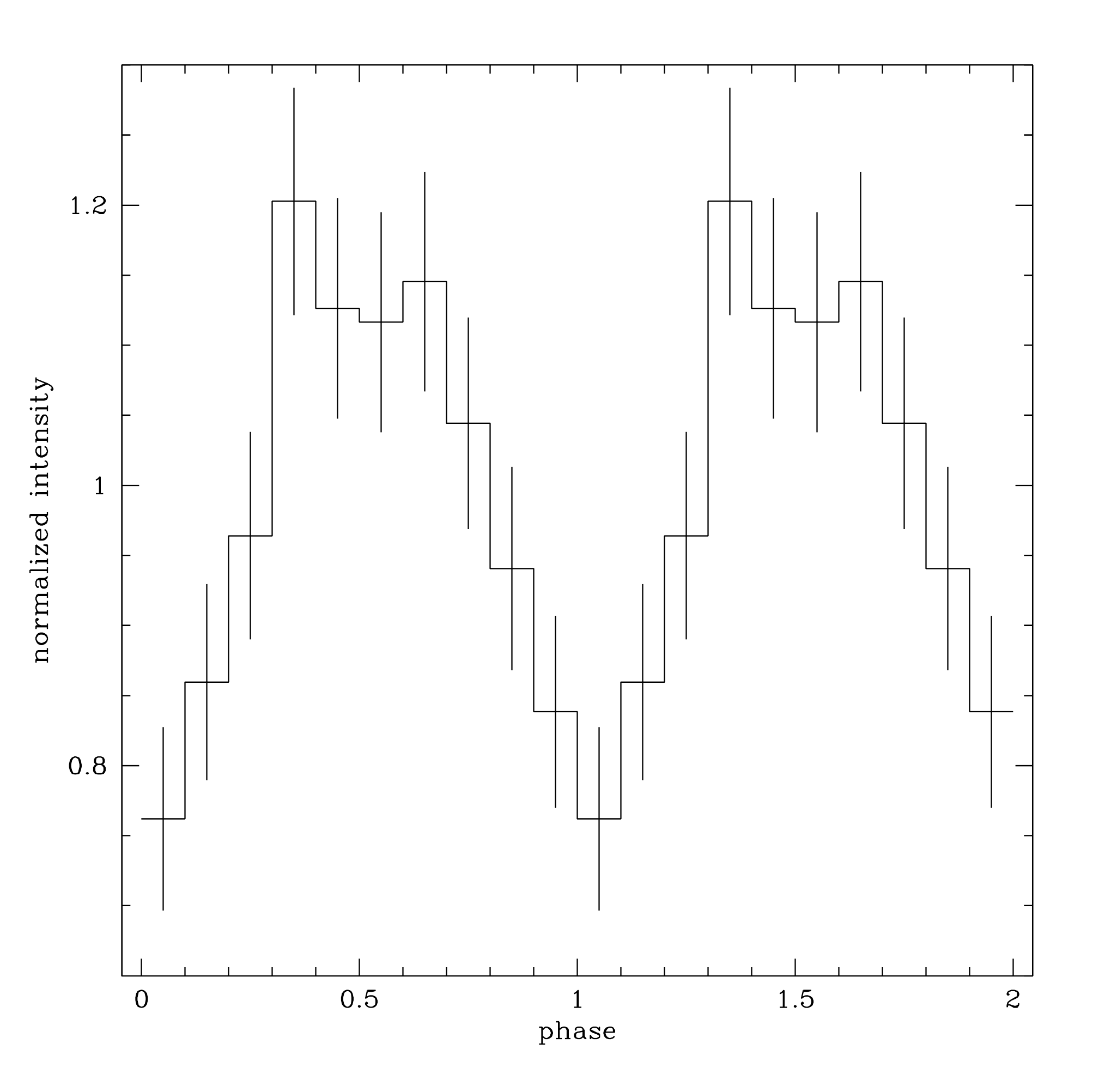}}
  }
  \mbox{
  \subfigure[ISGRI periodogram]{\includegraphics[width=0.5\textwidth]{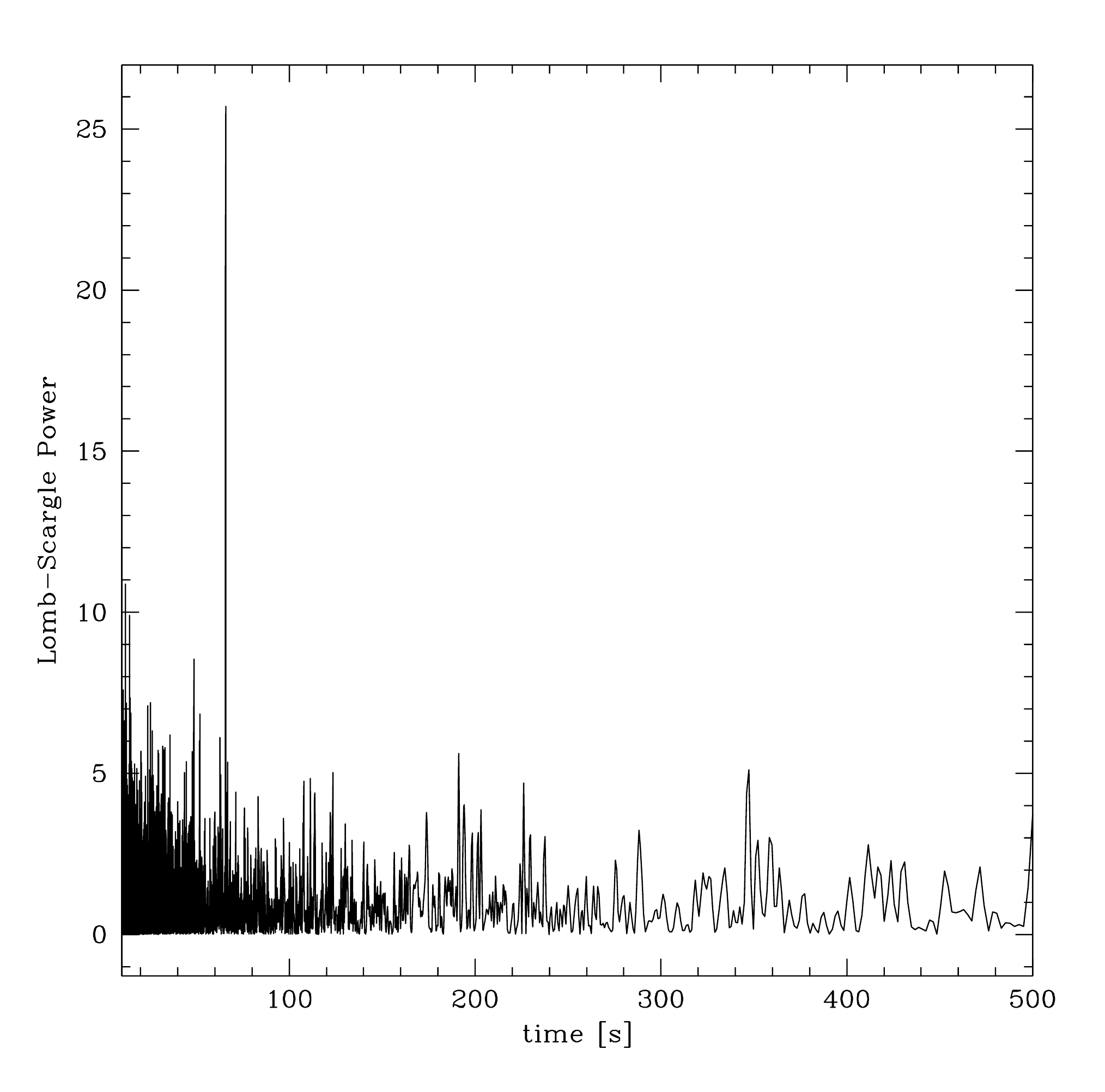}}
  \subfigure[ISGRI folded light curve, $P=65.789\pm0.009$~s and T0=MJD 52891.4]{\includegraphics[width=0.5\textwidth]{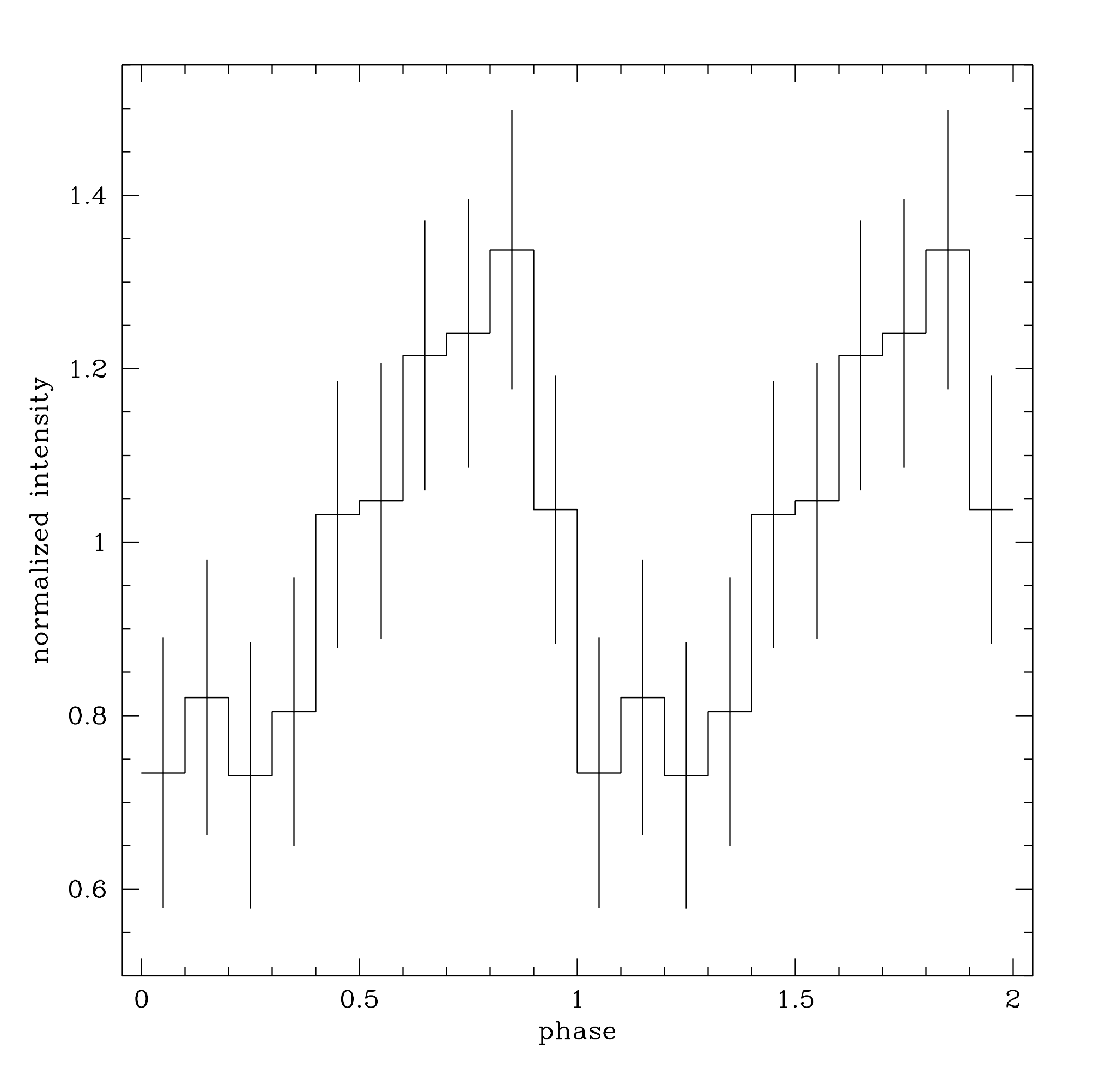}}
  }
  \caption{Periodograms and folded light curve for EPIC/pn ({\it top}, 0.4--10~keV) and ISGRI ({\it bottom}, 22-50~keV). The maximum power corresponds to $66.09\pm0.07$~s and $65.789\pm0.009$~s, respectively. The folded light curves use the derived periods and zero epochs of MJD 54190.391 and 52891.4.}
  \label{imaperiod}
\end{figure*}

We searched for the pulsation when computing the power density spectra of the EPIC light curve and of the 10~s-binned ISGRI light curve extracted during the long flare shown in Fig.~\ref{isgri_fl1}. We used the fast computing method of the Lomb-Scargle periodogram proposed by \citet{Pressal89}. The uncertainty of the period is computed using Eq.~14 in \citet{Horneal86}. Both periodograms are shown in Fig.~\ref{imaperiod} ({\it left column}). The maximum power is detected at $66.09\pm0.07$~s and $65.789\pm0.009$~s for pn and ISGRI, respectively. The corresponding folded light curves are also shown in Fig.~\ref{imaperiod} ({\it right column}) with zero epochs T0 of MJD 54190.391 and 52891.4. The pulse fractions, defined as $P_{\mathrm{f}}=(I_{\mathrm{max}}-I_{\mathrm{min}})/(I_{\mathrm{max}}+I_{\mathrm{min}})$ with $I_{\mathrm{max}}$ and $I_{\mathrm{min}}$ being the maximum and minimum of the intensities of the folded light curve, reach $22\pm6$ and $29\pm11$\%. The pulse profiles show one broad peak for pn and a slightly narrower one for ISGRI. As both pulsations are similar over a period of 3.6~yr, the timing analysis confirms that \integral\ and \xmm\ detected the same source, as it is unlikely that 2 X-ray sources with the same pulsation could be located in such a small box of a few arcseconds.

A pulsation of 132~s was also reported by \citet{Karaseval07,Karaseval08} using the same \xmm\ data presented here, along with the ISGRI data corresponding to the bright flare detected by \integral. No significant power appears at $\sim132$~s in both periodograms  in Fig.~\ref{imaperiod}. In the pn periodogram, the closest stronger signals appear at 123 and 139~s; yet their power are twice lower than the one at 66~s, and they do not appear in the ISGRI periodogram. Folding the light curve at twice the period we derived show a double-peak profile in both folded light curves. Still, both local peaks show similar shape and intensity. Therefore, the pulsation around $66$~s is likely to be the real pulsation, while the 132~s one is its first harmonic. \citet{Karaseval08} shows that no difference is observed in the pulse profile derived with ISGRI when folding the light curve with either 66 or 132~s. On the other hand, they show a slight difference between both pulse profiles derived with EPIC/pn, but this difference is not significant within the error bars.

\subsection{Spectral analysis}

\begin{figure}
\centering
\resizebox{\hsize}{!}{\includegraphics{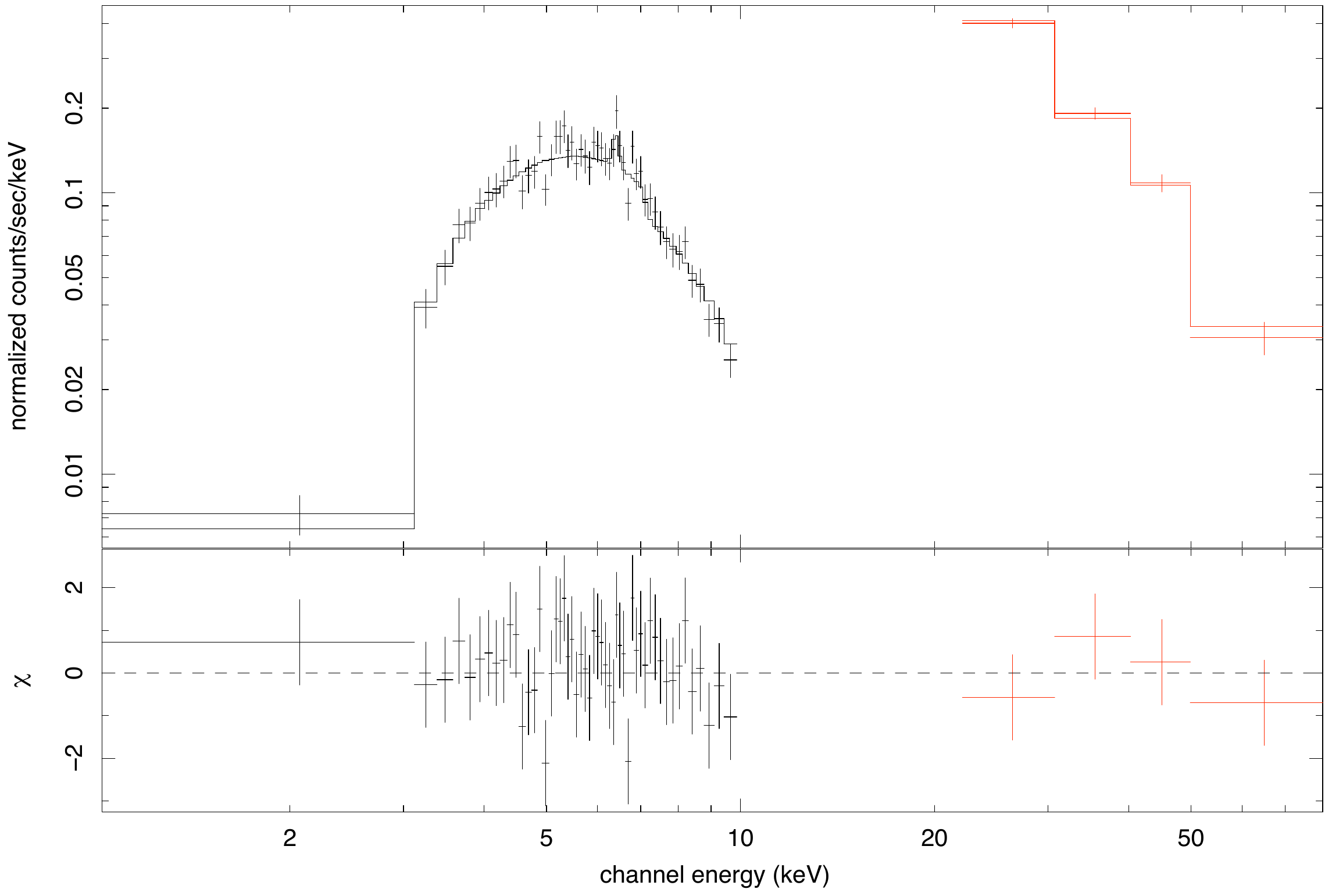}}
\caption{High-energy broad-band spectrum of \axj, combining the EPIC/pn and IBIS/ISGRI spectra. The best-fit model (absorbed PL with high-energy cutoff and an additional emission line) and residuals are also shown.}
\label{imaspec}
\end{figure}

The broad-band \xmm$-$\integral\ X-ray spectrum is shown in Fig.~\ref{imaspec}. The X-ray spectrum extracted with pn is combined with the hard X-ray one extracted during the bright flare seen in ISGRI. Therefore, an important constant factor must be applied, because the intensity difference of \axj\ between the 2 observations is large, and it may correspond to two different states of the source. Still, the combined spectrum will constrain the spectral model that will fit the continuum shape of the source. The X-ray spectral bins are grouped to possess at least 30 counts per channel, allowing use of the $\chi^{2}$ statistic.

First, the spectrum shows a strong absorption at low energies. Simple phenomenological models, such as absorbed black body (BB) or power-law (PL), fail to fit the data with \cnu$=$3.4 or 1.7 (each 106 d.o.f.), respectively. Disregarding the BB model, the absorbed PL clearly needs a high-energy cutoff to fit the data. The fit strongly improves with \cnu$=$1.03 (105 d.o.f.). We also explore the possibility of a line at 6.4~keV where some excess is visible. The tentative detection provides a good fit ($E_{\mathrm{line}}=6.41_{-0.07}^{+0.08}$~keV, $F_{\mathrm{line}}=(1.3\pm0.9)\times10^{-5}\ \mathrm{ph}\,\unit{cm}{-2}\,\unit{s}{-1}$, $EW_{\mathrm{line}}=52.5$~eV) of \cnu$=$0.99 (103 d.o.f.), but this is not significant because the line detection is $<3\sigma$. The possibility of an excess at low energies in addition to the absorption was also studied. However, the addition of a component at low energies is not significant as shown by the high value of the  F-test probability of 0.1. The best-fit parameters are listed in Table~\ref{tab_spec}.
\begin{table}
\caption{Best-fitting results for the broad-band spectrum.}
\label{tab_spec}      
\centering          
\begin{tabular}{l c l}
\hline\hline       
Parameters & Values & Unit \\
\hline                    
$\nh$ & $20.1_{-1.3}^{+1.5}$ & $10^{22}\ \mathrm{H}\,\unit{cm}{-2}$\\
$\Gamma$ & $1.0_{-0.3}^{+0.1}$ & ~\\
$\Ec$ & $21_{-3}^{+5}$ & keV \\
$\CI$ & 18 & ~\\
\cnu\ (dof) &1.03 & (105 d.o.f.)\\
unabs. 2--10~keV Flux & 0.2 & $10^{-10}\ \mathrm{ergs}\,\unit{cm}{-2}\,\unit{s}{-1}$\\ 
unabs. 22--50~keV Flux & 3.2 & $10^{-10}\ \mathrm{ergs}\,\unit{cm}{-2}\,\unit{s}{-1}$\\ 
unabs. 0.2--100~keV Flux & 6.9 & $10^{-10}\ \mathrm{ergs}\,\unit{cm}{-2}\,\unit{s}{-1}$\\ 
\hline
\end{tabular}
\begin{list}{}{}
\item Note: The model is {\tt cons*wabs*cutoffpl}. The unabsorbed flux is computed setting $\nh=0$.
\end{list}
\end{table}

\subsection{Identifying the optical/NIR counterpart}

\begin{figure}
\centering
\resizebox{\hsize}{!}{\includegraphics{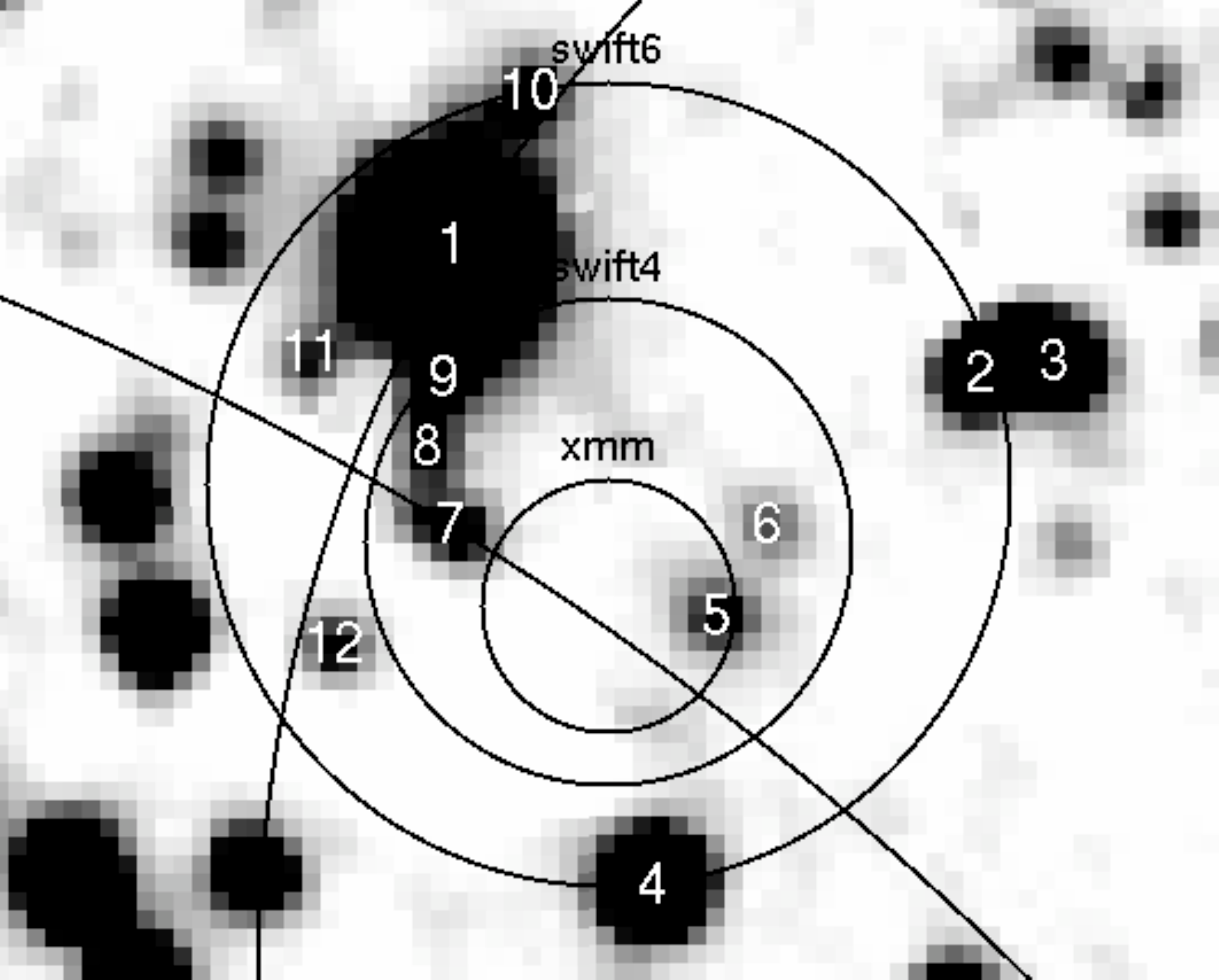}}
\caption{SOFI image in filter $J$. The 12 candidates within the large \swift\ error circle $(6\arcsec)$ are reported. Only candidate \#5 lies within the accurate X-ray position of \xmm\ $(2\arcsec)$.}
\label{imaJ}
\end{figure}
\begin{figure*}
\centering
\resizebox{\hsize}{!}{\includegraphics{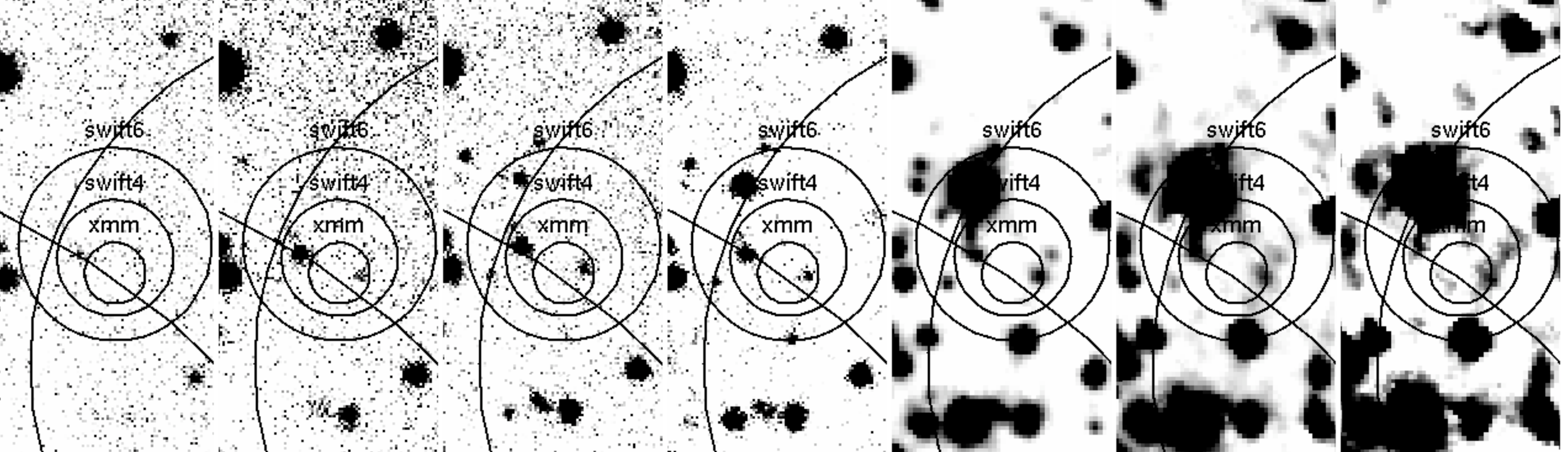}}
\caption{{\it Left to right}: images in filters: $V$, $R$, $I$, $Z$, $J$, $H$, and \Ks. The only candidate counterpart is  visible from filters $R$ to \Ks.}
\label{ima_optnir}
\end{figure*}

The field around \axj\ is very crowded, so it possesses several counterpart candidates within the error circles of the previously reported high-energy missions. In Fig.~\ref{imaJ}, we show the field in the $J$ band. With the higher accuracy of \swift, 3 2MASS candidates were reported \citep{Romanoal07a}. Still, with SOFI, we note that there are 11 candidates in the $J$ band within this $6\arcsec$ \swift\ box\footnote{The 3 2MASS candidates correspond to sources \#1, \#4 and a blended source constituted of \#2 and \#3.}. With the improved \swift\ position given by \citet{Kong07}, the 3 2MASS counterparts are ruled out, but 5 faint candidates remain. Taking the best X-ray position given by \xmm\ into account (see Sect.~\ref{secPos}), only candidate \#5 remains as a possible optical/NIR counterpart to \axj. The source was not detected in filters $U$, $B$, and $V$, but a faint source was detected in the other filters as highlighted in Fig.~\ref{ima_optnir}. The photometry of all sources is reported in Table~\ref{tab_photom}.
\begin{table*}
\caption{Optical and NIR photometry of the candidate counterparts located within the large \swift\ error circle.}
\label{tab_photom}      
\centering          
\begin{tabular}{c c c c c c}
\hline\hline       
Candidate$^\dagger$ & $R$ & $I$ & $J$ & $H$ & \Ks \\
\hline
\#1 & $22.6\pm0.5$ & $19.68\pm0.04$ & $12.49\pm0.01$ & $10.54\pm0.01$ & $9.18\pm0.01$ \\
\#2 & -- & -- & $16.32\pm0.06$ & $14.62\pm0.04$ & $13.51\pm0.04$ \\
\#3 & -- & -- & $15.84\pm0.04$ & $14.23\pm0.03$ & $13.27\pm0.05$ \\
\#4 & -- & -- & $15.27\pm0.01$ & $13.54\pm0.01$ & $12.45\pm0.02$ \\
\#5$^\ddagger$ & $21.3\pm0.1$ & $20.30\pm0.09$ & $17.42\pm0.03$ & $16.71\pm0.02$ & $15.75\pm0.07$ \\
\#6 & -- & -- & $18.08\pm0.03$ & $17.49\pm0.05$ & $16.1\pm0.1$ \\
\#7 & $20.15\pm0.05$ & $19.09\pm0.03$ & $17.33\pm0.05$ & $17.8\pm0.4$ & -- \\
\#8 & -- & -- & $17.88\pm0.09$ & $17.4\pm0.3$ & $15.6\pm0.1$ \\
\#9 & -- & -- & $17.43\pm0.06$ & $15.97\pm0.09$ & $14.5\pm0.2$ \\
\#10 & -- & $20.48\pm0.09$ & $17.23\pm0.04$ & -- & -- \\
\#11 & -- & -- & $18.11\pm0.07$ & $17.5\pm0.2$ & $16.0\pm0.1$ \\
\vspace{0.1cm}
\#12 & -- & $20.7\pm0.1$ & $17.84\pm0.07$ & $18.2\pm0.2$ & --  \\
\hline
$AM$ & 1.054 & 1.052 & 1.150 & 1.114 & 1.076 \\
$Z_{\mathrm{p}}$ & $-0.44\pm0.01$ & $0.39\pm0.01$ & $1.84\pm0.02$ & $2.01\pm0.02$ & $2.6\pm0.02$\\
$ext$ & $0.091$ & $0.051$ & $0.13\pm0.02$ & $0.10\pm0.02$ & $0.07\pm0.02$ \\
\hline
\end{tabular}
\begin{list}{}{}
\item[$^\dagger$]from Fig.~\ref{imaJ}.
\item[$^\ddagger$]located in the \xmm\ error circle.
\item Note:  At the bottom, we show the parameters for the photometric solutions: $AM$ the airmass, $Z_{\mathrm{p}}$ the zero-point and $ext$ the extinction.
\end{list}
\end{table*}

\section{Discussion and conclusion}\label{secDis}

The source possesses a long orbital period of $(185.5\pm1.1)/f$~d where $f=1,2,3,\ \mathrm{or}\ 4$ and a spin period of $\sim 66$~s. At the likely periastron passage, the source goes through an outburst whose duration is estimated to $\sim 12$~d (with an upper-limit of $\lesssim 18$~d). With such orbital period ($f=1$) and outburst duration, the source is bright $<10$\% of its time. The recurrence of such long outbursts related to the binary motion corresponds to type I outbursts observed in other BeXB \citep{Coe00, Negueruela07}. Furthermore, combining the observed pulsation and orbital period, \axj\ falls well in the BeXB arm of the Corbet diagram \citep[{\it i.e.} $P_{\mathrm{spin}}$ vs $P_{\mathrm{orb}}$;][]{Corbet84,Corbet86} strengthening its identification as a BeXB. The large orbital period rules out the possibility to associate this source with a low-mass X-ray binary whose companion is a star of the main sequence. Flaring activity in BeXB has never been observed except in EXO~2030$+$375 and recently in SWIFT~J1626.6$-$5156 \citep[][and references therein]{Reigal08}. For SWIFT~J1626.6$-$5156, the intensity varied by a factor 4 on a timescale of 450~s with sharp rise/decay of the count rate. This behaviour differs from the bright flare of \axj\ detected with ISGRI in revolution {\tt 110}, which shows a smooth rate's rise/decay, a variability factor of 8, and a duration of 0.5~d. This bright flare in revolution {\tt 110} is likely to be due to a sensitivity bias where only the peak flux of the outburst is significantly detected in individual pointings. However, flaring activity cannot be discarded for \axj\ because of the detections in pointings {\tt 01100038} and {\tt 01730024}, where the rate varies by a factor 3--4 on a timescale of 1~h in comparison to the average rates during the outbursts. Still, these detections may also be due to a sensitivity bias, because the count rates are similar to the 5.1$\sigma$ detection threshold.

A spin-down of \axj\ is observed between the 2 observations of the bright flare with \integral\ and the \xmm\ observation. We derived a spin-down of $\dot{P}=0.08\pm0.02\ \mathrm{s}\,\unit{yr}{-1}$ ({\it i.e.} $\dot{\nu}=(-6.2\pm1.4)\times 10^{-13}\ \mathrm{Hz}\,\unit{s}{-1}$). Such spin-down episodes have already been observed in other accreting pulsars, some examples being EXO~2030$+$375 with $\dot{\nu}=-3.4\times 10^{-14}\ \mathrm{Hz}\,\unit{s}{-1}$ and A0535$+$26 with $\dot{\nu}=-2.2\times 10^{-13}\ \mathrm{Hz}\,\unit{s}{-1}$ \citep[][and references therein]{Bildstenal97}. To have accretion onto an NS, the magnetic radius $R_{\mathrm{mag}}$ must be lower than the co-rotation radius $R_{\mathrm{cor}}$. Otherwise, the infalling matter is centrifugally expelled by the magnetic field \citep[the propeller regime,][]{Illarionoval75}. For an NS ($M=1.4\ \Ms$, $R=10$~km) with a spin period of 66~s, at the equilibrium point $R_{\mathrm{mag}}=R_{\mathrm{cor}}$, one estimates the magnetic field to be $B=6.5\times10^{12}\ \mathrm{G}\times (\dot{M}/10^{-10}\,\Ms\,\unit{yr}{-1})^{1/2}$. The magnetic radius depends on the accreted matter flow as $R_{\mathrm{mag}}\propto \dot{M}^{-2/7}$. Thus, the magnetic radius may exceed the co-rotation radius if the accretion rate becomes low enough, stopping the accretion. The infalling matter, expelled by the propeller effect, can then remove some angular momentum and decrease the spin period of the neutron star. With such a long orbital period, the NS is in fact usually far from the dense accreting zone, where such effects leading to a decrease in the spin period can occur.

The broad X- and hard X-ray spectrum of \axj\ is similar to other HMXB pulsars \citep{Whiteal95}. It is well-fitted with an absorbed power law and a high-energy cutoff with values $\nh=20.1_{-1.3}^{+1.5}\times10^{22}\ \unit{cm}{-2}$, $\Gamma=1.0_{-0.3}^{+0.1}$, and $\Ec=21_{-3}^{+5}$~keV. The large absorption is in good agreement with the previous observation \citep[\asca, \swift;][]{Sakanoal02,Romanoal07a,Kong07}; still, it is strongly constrained with the EPIC/pn data. This large absorption is an order of magnitude higher than the galactic absorption expected in the line of sight of \axj, $\nh(\mathrm{HI})=1.7\times10^{22}\ \unit{cm}{-2}$ \citep[interpolated from the HI map of][]{Dickeyal90}. Thus, the X-ray source is intrinsically absorbed. The combination of the EPIC/pn and IBIS/ISGRI spectra allowed us to better constrain the continuum getting a harder spectrum with a $\Gamma$ that decreases by a factor 2 in comparison with former estimates and the necessity of the high-energy cutoff. 

\begin{figure}
\centering
\resizebox{\hsize}{!}{\includegraphics{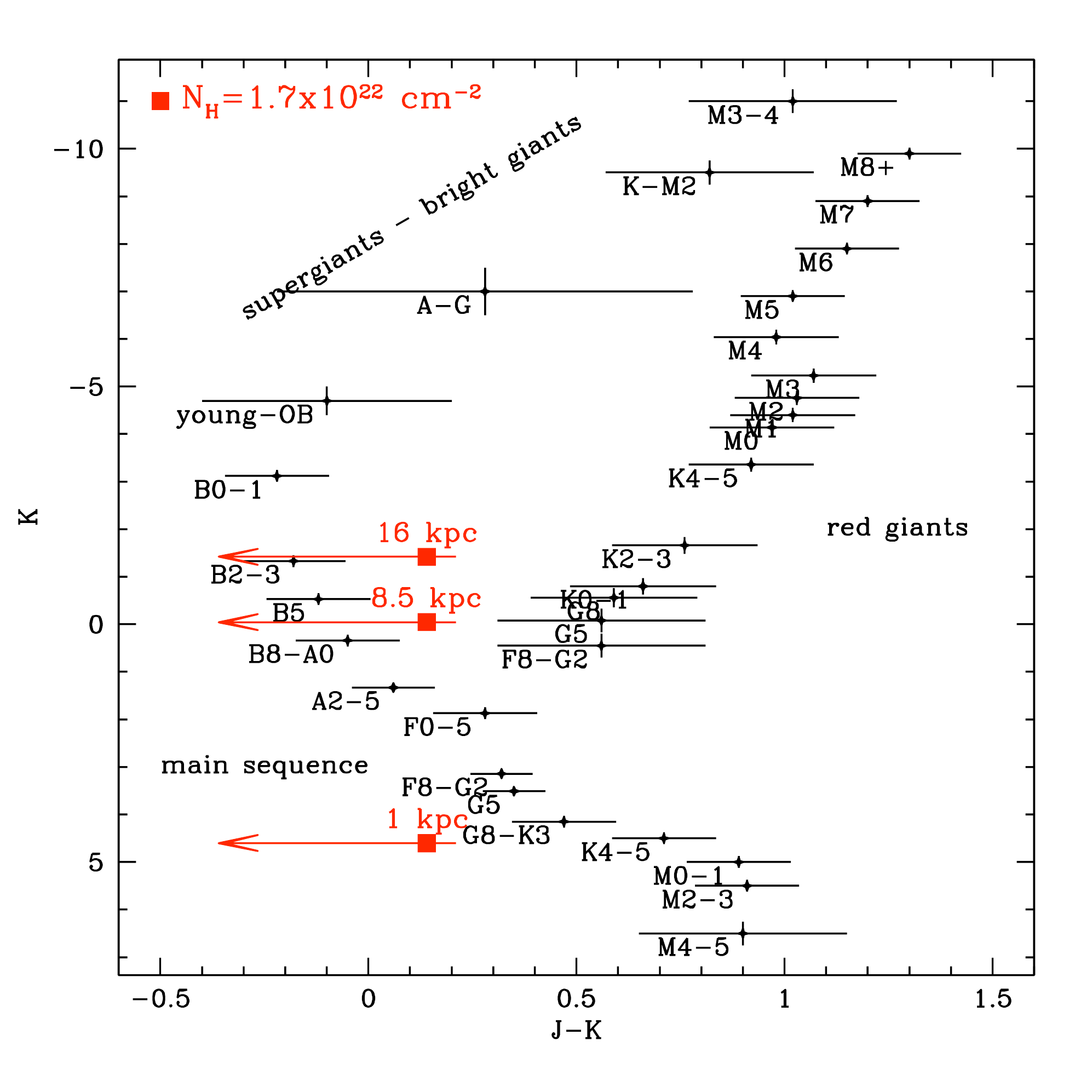}}
\caption{Colour-magnitude diagram. The magnitudes here correspond to absolute magnitude. The location of the source \#5 in the diagram is shown for  the galactic $\nh$ and several distances (1, 8.5 and 16 kpc). The arrows simulate the colour shift of $J-K\leq-0.5$ due to a disc-like ring of matter surrounding the star, assuming that it is a Be star.}
\label{ima_cmd}
\end{figure}
Only one candidate counterpart is located within the most accurate X-ray position of \xmm\ (see Sect.~\ref{secPos}). Using the relation of \citet{Predehlal95} that links the X-ray absorption and the optical extinction, $\nh\,(\unit{cm}{-2})/\Av\,(\mathrm{mag})=1.79\times 10^{21}$, the expected optical extinction $A_{\mathrm{V}}$ is 9.5~mag assuming the reported galactic $\nh=1.7\times10^{22}\ \unit{cm}{-2}$. Then, the NIR extinctions are estimated as $A_{\mathrm{J}}=2.680$,  $A_{\mathrm{H}}=1.663$ and $A_{\mathrm{K}}=1.064$ using $A_{\mathrm{J}}/\Av=0.282$,  $A_{\mathrm{H}}/\Av=0.175$ and $A_{\mathrm{K}}/\Av=0.112$ \citep{Riekeal85}. We derive the absolute magnitude as $M_{\mathrm{abs}}=m_{\mathrm{app}}-5\,log(\mathrm{distance}/10 \mathrm{pc})-\Av$. Assuming a distance of 8.5~kpc, we obtain $M_{\mathrm{J}}=0.09$, $M_{\mathrm{H}}=0.39$ and $M_{\mathrm{K}}=-0.05$\footnote{The \Ks\ magnitude is transformed into the $K$ magnitude using the relation $K-K_{\mathrm{s}}=-0.005(J-K)$ (SOFI user manual, p.7).}, so the colour is $(J-K)_{0}=0.14$. Thus, in a colour-magnitude diagram (CMD), the candidate companion star of \axj\ falls near the zone of the main sequence stars with spectral type between B5 and B8--A0 \seefig{ima_cmd}. In this case of a BeXB system, the disc-like ring of matter surrounding the B star can modify its colour $J-K$ by a large factor \citep[e.g. up to $-0.5$~mag in 4U~0115+63][]{Reigal07}. This is illustrated in Fig.~\ref{ima_cmd} by arrows of maximum length of 0.5 mag that strengthen the scenario of a B star located far away in the Galaxy, thus highly absorbed by the interstellar medium.
In another case where the star would be located nearer to us ($\sim 1$~kpc), thus with a lower extinction, the star would be a main sequence star of type KM. In that case, this star would probably not be the counterpart of \axj, because the large orbital period completely rules out the possibility of combining such KM star with this X-ray source. The possibility that the star is a red giant is also unlikely as it should combine a location far away in the Galaxy with low extinction. Spectroscopy of the candidate counterpart is needed to disentangle between the 2 possibilities, either a B star located far away in the Galaxy $\gtrsim8.5$~kpc and strongly absorbed or a normal star located nearer to us. Still, because of the X-ray properties observed for \axj\ and the absence of any other candidate, the B star is favoured. As the known BeXB have an earlier spectral type than B3 \citep{Negueruela98}, this would imply that the source is located at a very long distance of $>8.5$~kpc. Another argument in favour of the BeXB scenario is the reddening-free parameter $Q=(J-H)-1.70(H-K_{\mathrm{S}})$ \citep{Negueruelaal07a}. For candidate \#5, $Q\sim-0.9\pm0.1$, thus suggesting a high IR excess. Such excess is not expected in KM stars whose $Q$ values lie between 0.4--0.5; however, a factor $Q<0$ is expected for Be stars. Assuming a distance of 8.5~kpc, the 22--50~keV luminosity is $3\times10^{36}\ \es$ during the bright flare that occurred on MJD~52891 and $0.4-0.9\times10^{36}\ \es$ during the long outbursts, values typical of BeXB. The 0.2--10~keV luminosity is $0.2\times10^{36}\ \es$, a factor 2--4 lower than in \integral. We point out that, in the case of a BeXB with typical outbursts of the order of $\sim10^{36}\ \es$, and if the system was located nearer to us ($\lesssim 8$~kpc), \integral\ would detect such a source with a 22-50~keV count rate $\gtrsim 2$~\cps, which is not the case.

In the case of a BeXB, \axj\ shows a high intrinsic absorption of $\nh=20.1_{-1.3}^{+1.5}\times10^{22}\ \unit{cm}{-2}$ that is quite uncommon in other BeXB. Indeed, most of the BeXB have density columns closer to the galactic absorption with $\nh\sim 1-3\times N_{\mathrm{H}} ^{\mathrm{Gal}}$ \citep[see Fig.~15 of][]{Bodagheeal07}. Only one BeXB exhibits such high absorption: 2S~1845$-$024. The properties of this source are very similar to \axj\ with $P_{\mathrm{orb}}=242.18\pm0.01$~d, outburst duration of $\sim 13$~d, $P_{\mathrm{spin}}=94.9$~s (with episodes of spin-up and spin-down) and $\nh=25\pm10\times10^{22}\ \unit{cm}{-2}$ \citep[][and references therein]{Koyamaal90,Soffittaal98,Fingeral99}. They classified 2S~1845$-$024 as a BeXB (located at 10~kpc\footnote{This value can be considered as an upper limit as they derive it taking the minimum observed $\nh$ as interstellar absorption without taking the intrinsic absorption into account.}) from its X-ray properties; however, no counterpart has been identified yet. Thus, \axj\ might be the first identified counterpart BeXB with high intrinsic absorption similar to the obscured SGXB revealed by \integral. Spectral properties of pulsars, either SGXB or BeXB, are very similar \citep[see e.g.][]{Whiteal95}. As IBIS/ISGRI is not affected by absorption, \integral\ allowed a large population of obscured SGXB to be revealed with $\nh$ values of $\gtrsim10^{23}\ \unit{cm}{-2}$. That such highly-absorbed BeXB have not been observed yet \citep{Bodagheeal07} is perhaps not so surprising since BeXB are mostly transient sources, and the sources must also be observed during the outbursts by another X-ray emission with high sensitivity at low energies (such as \xmm\ or \chandra) in order to constrain the absorption. Such a population of obscured BeXB might still have to be revealed. Indeed, the only BeXB among the \integral\ sources reported in Fig.~15 of \citet{Bodagheeal07} is the 3rd highly-absorbed BeXB (after \axj\ and 2S~1845$-$024).

All the X-ray properties observed in \axj\ lead to classifing this object as an X-ray binary, most probably an HMXB with a Be companion star whose compact object is an NS. The only optical/NIR candidate counterpart located inside the best X-ray error circle is compatible with a B star located far inside the galaxy and suffering large extinction. Besides, the lack of a supergiant companion and the duration/smoothness of the outbursts rule out the classification of this source as an SFXT, as proposed in \citet{Zand05} and \citet{Sgueraal06}. Instead, \axj\ seems to be an obscured BeXB located far away in the Galaxy.

\begin{acknowledgements}
The authors warmly thank the \xmm\ team for accepting the ToO request. SC thanks the ESO staff, and especially Valenti Ivanov, for performing the ToO observations.  We also warmly thank the referee I.~Negueruela for his useful comments that improved the manuscript. We also thank J\'er\^ome Rodriguez for a careful re-reading of the manuscript and for giving useful comments. We would like to thank A. Lutovinov for discussing the nature of this source at the 5th \integral\ Anniversary workshop.
\end{acknowledgements}

\bibliographystyle{aa}
\bibliography{../paper/biblio}

\begin{thebibliography}{50}
\expandafter\ifx\csname natexlab\endcsname\relax\def\natexlab#1{#1}\fi

\bibitem[{{Arnaud}(1996)}]{Arnaud96}
{Arnaud}, K.~A. 1996, in Astronomical Society of the Pacific Conference Series,
  Vol. 101, Astronomical Data Analysis Software and Systems V, ed. G.~H.
  {Jacoby} \& J.~{Barnes}, 17--+

\bibitem[{{Bildsten} {et~al.}(1997){Bildsten}, {Chakrabarty}, {Chiu}, {Finger},
  {Koh}, {Nelson}, {Prince}, {Rubin}, {Scott}, {Stollberg}, {Vaughan},
  {Wilson}, \& {Wilson}}]{Bildstenal97}
{Bildsten}, L., {Chakrabarty}, D., {Chiu}, J., {et~al.} 1997, \apjs, 113, 367

\bibitem[{{Bird} {et~al.}(2007){Bird}, {Malizia}, {Bazzano}, {Barlow},
  {Bassani}, {Hill}, {B{\'e}langer}, {Capitanio}, {Clark}, {Dean}, {Fiocchi},
  {G{\"o}tz}, {Lebrun}, {Molina}, {Produit}, {Renaud}, {Sguera}, {Stephen},
  {Terrier}, {Ubertini}, {Walter}, {Winkler}, \& {Zurita}}]{Birdal07}
{Bird}, A.~J., {Malizia}, A., {Bazzano}, A., {et~al.} 2007, \apjs, 170, 175

\bibitem[{{Bodaghee} {et~al.}(2007){Bodaghee}, {Courvoisier}, {Rodriguez},
  {Beckmann}, {Produit}, {Hannikainen}, {Kuulkers}, {Willis}, \&
  {Wendt}}]{Bodagheeal07}
{Bodaghee}, A., {Courvoisier}, T.~J.-L., {Rodriguez}, J., {et~al.} 2007, \aap,
  467, 585

\bibitem[{{Coe}(2000)}]{Coe00}
{Coe}, M.~J. 2000, in Astronomical Society of the Pacific Conference Series,
  Vol. 214, IAU Colloq. 175: The Be Phenomenon in Early-Type Stars, ed. M.~A.
  {Smith}, H.~F. {Henrichs}, \& J.~{Fabregat}, 656--+

\bibitem[{{Corbet}(1984)}]{Corbet84}
{Corbet}, R.~H.~D. 1984, \aap, 141, 91

\bibitem[{{Corbet}(1986)}]{Corbet86}
{Corbet}, R.~H.~D. 1986, \mnras, 220, 1047

\bibitem[{{Courvoisier} {et~al.}(2003){Courvoisier}, {Walter}, {Beckmann},
  {Dean}, {Dubath}, {Hudec}, {Kretschmar}, {Mereghetti}, {Montmerle},
  {Mowlavi}, {Paltani}, {Preite Martinez}, {Produit}, {Staubert}, {Strong},
  {Swings}, {Westergaard}, {White}, {Winkler}, \&
  {Zdziarski}}]{Courvoisieral03}
{Courvoisier}, T.~J.-L., {Walter}, R., {Beckmann}, V., {et~al.} 2003, \aap,
  411, L53

\bibitem[{{Dickey} \& {Lockman}(1990)}]{Dickeyal90}
{Dickey}, J.~M. \& {Lockman}, F.~J. 1990, \araa, 28, 215

\bibitem[{{Finger} {et~al.}(1999){Finger}, {Bildsten}, {Chakrabarty}, {Prince},
  {Scott}, {Wilson}, {Wilson}, \& {Zhang}}]{Fingeral99}
{Finger}, M.~H., {Bildsten}, L., {Chakrabarty}, D., {et~al.} 1999, \apj, 517,
  449

\bibitem[{{G{\"o}tz} {et~al.}(2007){G{\"o}tz}, {Falanga}, {Senziani}, {De
  Luca}, {Schanne}, \& {von Kienlin}}]{Gotzal07}
{G{\"o}tz}, D., {Falanga}, M., {Senziani}, F., {et~al.} 2007, \apjl, 655, L101

\bibitem[{{Grebenev} \& {Sunyaev}(2007)}]{Grebeneval07}
{Grebenev}, S.~A. \& {Sunyaev}, R.~A. 2007, Astronomy Letters, 33, 149

\bibitem[{{Gros} {et~al.}(2003){Gros}, {Goldwurm}, {Cadolle-Bel}, {Goldoni},
  {Rodriguez}, {Foschini}, {Del Santo}, \& {Blay}}]{Grosal03}
{Gros}, A., {Goldwurm}, A., {Cadolle-Bel}, M., {et~al.} 2003, \aap, 411, L179

\bibitem[{{Horne} \& {Baliunas}(1986)}]{Horneal86}
{Horne}, J.~H. \& {Baliunas}, S.~L. 1986, \apj, 302, 757

\bibitem[{{Illarionov} \& {Sunyaev}(1975)}]{Illarionoval75}
{Illarionov}, A.~F. \& {Sunyaev}, R.~A. 1975, \aap, 39, 185

\bibitem[{{in't Zand}(2005)}]{Zand05}
{in't Zand}, J.~J.~M. 2005, \aap, 441, L1

\bibitem[{{Jansen} {et~al.}(2001){Jansen}, {Lumb}, {Altieri}, {Clavel}, {Ehle},
  {Erd}, {Gabriel}, {Guainazzi}, {Gondoin}, {Much}, {Munoz}, {Santos},
  {Schartel}, {Texier}, \& {Vacanti}}]{Jansenal01}
{Jansen}, F., {Lumb}, D., {Altieri}, B., {et~al.} 2001, \aap, 365, L1

\bibitem[{{Karasev} {et~al.}(2007){Karasev}, {Tsygankov}, {Lutovinov},
  {Churazov}, \& {Sunyaev}}]{Karaseval07}
{Karasev}, D., {Tsygankov}, S., {Lutovinov}, A., {Churazov}, E., \& {Sunyaev},
  R. 2007, The Astronomer's Telegram, 1245, 1

\bibitem[{{Karasev} {et~al.}(2008){Karasev}, {Tsygankov}, \&
  {Lutovinov}}]{Karaseval08}
{Karasev}, D.~I., {Tsygankov}, S.~S., \& {Lutovinov}, A.~A. 2008, \mnras, L50

\bibitem[{{Kong}(2007)}]{Kong07}
{Kong}, A.~K.~H. 2007, The Astronomer's Telegram, 1039, 1

\bibitem[{{Koyama} {et~al.}(1990){Koyama}, {Kunieda}, {Takeuchi}, \&
  {Tawara}}]{Koyamaal90}
{Koyama}, K., {Kunieda}, H., {Takeuchi}, Y., \& {Tawara}, Y. 1990, \pasj, 42,
  L59

\bibitem[{{Kuulkers} {et~al.}(2007){Kuulkers}, {Shaw}, {Paizis}, {Chenevez},
  {Brandt}, {Courvoisier}, {Domingo}, {Ebisawa}, {Kretschmar}, {Markwardt},
  {Mowlavi}, {Oosterbroek}, {Orr}, {R{\'{\i}}squez}, {Sanchez-Fernandez}, \&
  {Wijnands}}]{Kuulkersal07}
{Kuulkers}, E., {Shaw}, S.~E., {Paizis}, A., {et~al.} 2007, \aap, 466, 595

\bibitem[{{Landolt}(1992)}]{Landolt92}
{Landolt}, A.~U. 1992, \aj, 104, 340

\bibitem[{{Lebrun} {et~al.}(2003){Lebrun}, {Leray}, {Lavocat}, {Cr{\'e}tolle},
  {Arqu{\`e}s}, {Blondel}, {Bonnin}, {Bou{\`e}re}, {Cara}, {Chaleil}, {Daly},
  {Desages}, {Dzitko}, {Horeau}, {Laurent}, {Limousin}, {Mathy}, {Mauguen},
  {Meignier}, {Molini{\'e}}, {Poindron}, {Rouger}, {Sauvageon}, \&
  {Tourrette}}]{Lebrunal03}
{Lebrun}, F., {Leray}, J.~P., {Lavocat}, P., {et~al.} 2003, \aap, 411, L141

\bibitem[{{Leyder} {et~al.}(2007){Leyder}, {Walter}, {Lazos}, {Masetti}, \&
  {Produit}}]{Leyderal07}
{Leyder}, J.-C., {Walter}, R., {Lazos}, M., {Masetti}, N., \& {Produit}, N.
  2007, \aap, 465, L35

\bibitem[{{Liu} {et~al.}(2000){Liu}, {van Paradijs}, \& {van den
  Heuvel}}]{Liual00}
{Liu}, Q.~Z., {van Paradijs}, J., \& {van den Heuvel}, E.~P.~J. 2000, \aaps,
  147, 25

\bibitem[{{Liu} {et~al.}(2006){Liu}, {van Paradijs}, \& {van den
  Heuvel}}]{Liual06}
{Liu}, Q.~Z., {van Paradijs}, J., \& {van den Heuvel}, E.~P.~J. 2006, \aap,
  455, 1165

\bibitem[{{Negueruela}(1998)}]{Negueruela98}
{Negueruela}, I. 1998, \aap, 338, 505

\bibitem[{{Negueruela}(2007)}]{Negueruela07}
{Negueruela}, I. 2007, in Astronomical Society of the Pacific Conference
  Series, Vol. 367, Massive Stars in Interactive Binaries, ed. N.~{St.-Louis}
  \& A.~F.~J. {Moffat}, 477--+

\bibitem[{{Negueruela} \& {Schurch}(2007)}]{Negueruelaal07a}
{Negueruela}, I. \& {Schurch}, M.~P.~E. 2007, \aap, 461, 631

\bibitem[{{Negueruela} {et~al.}(2006){Negueruela}, {Smith}, {Reig}, {Chaty}, \&
  {Torrej{\'o}n}}]{Negueruelaal06}
{Negueruela}, I., {Smith}, D.~M., {Reig}, P., {Chaty}, S., \& {Torrej{\'o}n},
  J.~M. 2006, in ESA SP-604: The X-ray Universe 2005, ed. A.~{Wilson}, 165--170

\bibitem[{{Pellizza} {et~al.}(2006){Pellizza}, {Chaty}, \&
  {Negueruela}}]{Pellizzaal06}
{Pellizza}, L.~J., {Chaty}, S., \& {Negueruela}, I. 2006, \aap, 455, 653

\bibitem[{{Persson} {et~al.}(1998){Persson}, {Murphy}, {Krzeminski}, {Roth}, \&
  {Rieke}}]{Perssonal98}
{Persson}, S.~E., {Murphy}, D.~C., {Krzeminski}, W., {Roth}, M., \& {Rieke},
  M.~J. 1998, \aj, 116, 2475

\bibitem[{{Predehl} \& {Schmitt}(1995)}]{Predehlal95}
{Predehl}, P. \& {Schmitt}, J.~H.~M.~M. 1995, \aap, 293, 889

\bibitem[{{Press} \& {Rybicki}(1989)}]{Pressal89}
{Press}, W.~H. \& {Rybicki}, G.~B. 1989, \apj, 338, 277

\bibitem[{{Psaltis}(2006)}]{Psaltis06}
{Psaltis}, D. 2006, {Accreting neutron stars and black holes: a decade of
  discoveries} (Compact stellar X-ray sources), 1--38

\bibitem[{{Reig} {et~al.}(2008){Reig}, {Belloni}, {Israel}, {Campana},
  {Gehrels}, \& {Homan}}]{Reigal08}
{Reig}, P., {Belloni}, T., {Israel}, G.~L., {et~al.} 2008, ArXiv e-prints, 804

\bibitem[{{Reig} {et~al.}(2007){Reig}, {Larionov}, {Negueruela}, {Arkharov}, \&
  {Kudryavtseva}}]{Reigal07}
{Reig}, P., {Larionov}, V., {Negueruela}, I., {Arkharov}, A.~A., \&
  {Kudryavtseva}, N.~A. 2007, \aap, 462, 1081

\bibitem[{{Rieke} \& {Lebofsky}(1985)}]{Riekeal85}
{Rieke}, G.~H. \& {Lebofsky}, M.~J. 1985, \apj, 288, 618

\bibitem[{{Romano} {et~al.}(2007){Romano}, {Mangano}, {Cucchiara}, \&
  {Sidoli}}]{Romanoal07a}
{Romano}, P., {Mangano}, V., {Cucchiara}, A., \& {Sidoli}, L. 2007, The
  Astronomer's Telegram, 1040, 1

\bibitem[{{Sakano} {et~al.}(2002){Sakano}, {Koyama}, {Murakami}, {Maeda}, \&
  {Yamauchi}}]{Sakanoal02}
{Sakano}, M., {Koyama}, K., {Murakami}, H., {Maeda}, Y., \& {Yamauchi}, S.
  2002, \apjs, 138, 19

\bibitem[{{Sguera} {et~al.}(2006){Sguera}, {Bazzano}, {Bird}, {Dean},
  {Ubertini}, {Barlow}, {Bassani}, {Clark}, {Hill}, {Malizia}, {Molina}, \&
  {Stephen}}]{Sgueraal06}
{Sguera}, V., {Bazzano}, A., {Bird}, A.~J., {et~al.} 2006, \apj, 646, 452

\bibitem[{{Smith} {et~al.}(2006){Smith}, {Heindl}, {Markwardt}, {Swank},
  {Negueruela}, {Harrison}, \& {Huss}}]{Smithal06}
{Smith}, D.~M., {Heindl}, W.~A., {Markwardt}, C.~B., {et~al.} 2006, \apj, 638,
  974

\bibitem[{{Soffitta} {et~al.}(1998){Soffitta}, {Tomsick}, {Harmon}, {Costa},
  {Ford}, {Tavani}, {Zhang}, \& {Kaaret}}]{Soffittaal98}
{Soffitta}, P., {Tomsick}, J.~A., {Harmon}, B.~A., {et~al.} 1998, \apjl, 494,
  L203+

\bibitem[{{Str{\"u}der} {et~al.}(2001){Str{\"u}der}, {Briel}, {Dennerl},
  {Hartmann}, {Kendziorra}, {Meidinger}, {Pfeffermann}, {Reppin}, {Aschenbach},
  {Bornemann}, {Br{\"a}uninger}, {Burkert}, {Elender}, {Freyberg}, {Haberl},
  {Hartner}, {Heuschmann}, {Hippmann}, {Kastelic}, {Kemmer}, {Kettenring},
  {Kink}, {Krause}, {M{\"u}ller}, {Oppitz}, {Pietsch}, {Popp}, {Predehl},
  {Read}, {Stephan}, {St{\"o}tter}, {Tr{\"u}mper}, {Holl}, {Kemmer}, {Soltau},
  {St{\"o}tter}, {Weber}, {Weichert}, {von Zanthier}, {Carathanassis}, {Lutz},
  {Richter}, {Solc}, {B{\"o}ttcher}, {Kuster}, {Staubert}, {Abbey}, {Holland},
  {Turner}, {Balasini}, {Bignami}, {La Palombara}, {Villa}, {Buttler},
  {Gianini}, {Lain{\'e}}, {Lumb}, \& {Dhez}}]{Struderal01}
{Str{\"u}der}, L., {Briel}, U., {Dennerl}, K., {et~al.} 2001, \aap, 365, L18

\bibitem[{{Turner} {et~al.}(2001){Turner}, {Abbey}, {Arnaud}, {Balasini},
  {Barbera}, {Belsole}, {Bennie}, {Bernard}, {Bignami}, {Boer}, {Briel},
  {Butler}, {Cara}, {Chabaud}, {Cole}, {Collura}, {Conte}, {Cros}, {Denby},
  {Dhez}, {Di Coco}, {Dowson}, {Ferrando}, {Ghizzardi}, {Gianotti}, {Goodall},
  {Gretton}, {Griffiths}, {Hainaut}, {Hochedez}, {Holland}, {Jourdain},
  {Kendziorra}, {Lagostina}, {Laine}, {La Palombara}, {Lortholary}, {Lumb},
  {Marty}, {Molendi}, {Pigot}, {Poindron}, {Pounds}, {Reeves}, {Reppin},
  {Rothenflug}, {Salvetat}, {Sauvageot}, {Schmitt}, {Sembay}, {Short},
  {Spragg}, {Stephen}, {Str{\"u}der}, {Tiengo}, {Trifoglio}, {Tr{\"u}mper},
  {Vercellone}, {Vigroux}, {Villa}, {Ward}, {Whitehead}, \&
  {Zonca}}]{Turneral01}
{Turner}, M.~J.~L., {Abbey}, A., {Arnaud}, M., {et~al.} 2001, \aap, 365, L27

\bibitem[{{Ubertini} {et~al.}(2003){Ubertini}, {Lebrun}, {Di Cocco}, {Bazzano},
  {Bird}, {Broenstad}, {Goldwurm}, {La Rosa}, {Labanti}, {Laurent}, {Mirabel},
  {Quadrini}, {Ramsey}, {Reglero}, {Sabau}, {Sacco}, {Staubert}, {Vigroux},
  {Weisskopf}, \& {Zdziarski}}]{Ubertinial03}
{Ubertini}, P., {Lebrun}, F., {Di Cocco}, G., {et~al.} 2003, \aap, 411, L131

\bibitem[{{White} {et~al.}(1995){White}, {Nagase}, \& {Parmar}}]{Whiteal95}
{White}, N.~E., {Nagase}, F., \& {Parmar}, A.~N. 1995, {X-ray binaries}
  (Cambridge Astrophysics Series, Cambridge, MA: Cambridge University Press,
  |c1995, edited by Lewin, Walter H.G.; Van Paradijs, Jan; Van den Heuvel,
  Edward P.J.), 1--57

\bibitem[{{Winkler} {et~al.}(2003){Winkler}, {Courvoisier}, {Di Cocco},
  {Gehrels}, {Gim{\'e}nez}, {Grebenev}, {Hermsen}, {Mas-Hesse}, {Lebrun},
  {Lund}, {Palumbo}, {Paul}, {Roques}, {Schnopper}, {Sch{\"o}nfelder},
  {Sunyaev}, {Teegarden}, {Ubertini}, {Vedrenne}, \& {Dean}}]{Winkleral03}
{Winkler}, C., {Courvoisier}, T.~J.-L., {Di Cocco}, G., {et~al.} 2003, \aap,
  411, L1

\bibitem[{{Zurita Heras} {et~al.}(2007){Zurita Heras}, {Chaty}, \&
  {Rodriguez}}]{Zuritaal07}
{Zurita Heras}, J.~A., {Chaty}, S., \& {Rodriguez}, J. 2007, The Astronomer's
  Telegram, 1035, 1

\end{thebibliography}

\end{document}